\documentclass[acmsmall]{acmart}

\renewcommand\footnotetextcopyrightpermission[1]{}
\usepackage{array}
\usepackage{booktabs}
\usepackage{graphicx}
\usepackage[flushleft]{threeparttable}
\newcolumntype{P}[1]{>{\centering\arraybackslash}p{#1}}
\newcolumntype{M}[1]{>{\centering\arraybackslash}m{#1}}
\usepackage{makecell}
\usepackage{listings}
\usepackage{xcolor}
\usepackage[caption=false,font=footnotesize,farskip=0pt,captionskip=0pt]{subfig}
\usepackage{color}

\AtBeginDocument{%
  }

\begin{document}

\title{A Step-by-Step Guide to Creating a Robust Autonomous Drone Testing Pipeline}


\author{Yupeng Jiang}
\email{yupeng.jiang@hdr.mq.edu.au}
\author{Yao Deng}
\email{yao.deng@mq.edu.au}
\author{Sebastian Schroder}
\email{sebastian.schroder@mq.edu.au}
\author{Linfeng Liang}
\email{linfeng.liang@hdr.mq.edu.au}
\author{Suhaas Gambhir}
\email{suhaas.gambhir@mq.edu.au}
\author{Alice James}
\email{alice.james@mq.edu.au}
\author{Avishkar Seth}
\email{avishkar.seth@mq.edu.au}
\author{James Pirrie}
\email{james.pirrie@students.mq.edu.au}
\author{Yihao Zhang}
\email{yihao.zhang@mq.edu.au}
\author{Xi Zheng}
\email{james.zheng@mq.edu.au}
\affiliation{%
  \institution{Macquarie University}
  \city{Sydney}
  \country{Australia}
}

\renewcommand{\shortauthors}{Jiang et al.}

\begin{abstract}
Autonomous drones are rapidly reshaping industries ranging from aerial delivery and infrastructure inspection to environmental monitoring and disaster response. Ensuring the safety, reliability, and efficiency of these systems is paramount as they transition from research prototypes to mission-critical platforms. This paper presents a step-by-step guide to establishing a robust autonomous drone testing pipeline, covering each critical stage: Software-in-the-Loop (SIL) Simulation Testing, Hardware-in-the-Loop (HIL) Testing, Controlled Real-World Testing, and In-Field Testing. Using practical examples, including the marker-based autonomous landing system, we demonstrate how to systematically verify drone system behaviors, identify integration issues, and optimize performance. Furthermore, we highlight emerging trends shaping the future of drone testing, including the integration of Neurosymbolic and LLMs, creating co-simulation environments, and Digital Twin-enabled simulation-based testing techniques. By following this pipeline, developers and researchers can achieve comprehensive validation, minimize deployment risks, and prepare autonomous drones for safe and reliable real-world operations.
\end{abstract}





\maketitle

\section{Introduction}
Unmanned Aerial Vehicles (UAVs), commonly known as drones, have rapidly transitioned from niche research prototypes to indispensable tools across a wide range of industries. From precision agriculture \cite{GreyB} and aerial photography \cite{SDW} to emergency response \cite{Wired} and logistics \cite{DHL}, drones are revolutionizing the way tasks are performed in complex and dynamic environments. Among these, autonomous drones—capable of perceiving, reasoning, and making decisions without direct human control—represent the frontier of aerial robotics.

As autonomous drones become more prevalent and their applications more safety-critical, ensuring the reliability, safety, and efficiency of their operations has emerged as a crucial concern. Unlike manually piloted drones, autonomous systems must interpret sensor data, localize themselves in potentially GPS-denied environments, plan paths, avoid obstacles, and control flight dynamics while operating under uncertain and rapidly changing conditions \cite{mittal2019vision}.

To manage this complexity, modern autonomous drone systems are typically organized as multi-module architectures, where each functional component (e.g., perception, localization, planning, control) operates semi-independently and interacts through well-defined interfaces \cite{liang2025garl}. This design paradigm, often built on middleware platforms such as the Robot Operating System (ROS) \cite{quigley2009ros}, mirrors practices in broader autonomous driving systems (ADS) \cite{deng2022scenario}. Indeed, the majority of ADS practitioners report working on modular systems rather than monolithic end-to-end models, citing scalability, debuggability, and safety as primary motivators \cite{Lou2021TestingOA}.

While modularity enhances flexibility and maintainability, it also introduces new challenges in system validation and verification. Robust testing processes are essential to ensure that individual modules, as well as their complex interactions, function correctly under both normal and edge-case scenarios. System failures in autonomous drones may result in significant consequences, including property damage, financial losses, or threats to human safety \cite{shakhatreh2019unmanned}. Therefore, the necessity for a systematic and comprehensive testing pipeline has never been greater.

This work aims to address this pressing need by providing a structured, step-by-step guide to designing and implementing a robust testing pipeline for autonomous drones. From simulation environments and software-in-the-loop (SIL) testing to hardware-in-the-loop (HIL) validation and real-world trials, we will introduce best practices, tools, and methodologies that ensure safe and reliable drone operations. Throughout this paper, real-world examples and case studies such as the marker-based autonomous landing system \cite{liang2025garl} will be used to demonstrate practical applications and highlight challenges faced at each stage of the testing process.

\subsection{Popular Autonomous Drone Systems} \label{multi-module}
Modern autonomous drones are complex cyber-physical systems that operate with minimal or no human intervention across a wide range of missions, including aerial surveying, package delivery, search and rescue, and infrastructure inspection \cite{sebbane2015smart}. Rather than being built as monolithic AI models, most real-world drone platforms adopt a modular system architecture, often structured around the Robot Operating System (ROS) \cite{quigley2009ros}. In such systems, functional components such as perception, localization, planning, and control operate as discrete ROS nodes that communicate through standardized message-passing interfaces \cite{shakhatreh2019unmanned, gustave2020functional, meier2015px4, causa2021multiple, antonopoulos2022ros, aliane2024survey}. This architecture enables the integration of both machine learning-based modules (e.g., visual object or marker detection) and rule-based control libraries (e.g., PX4 \cite{meier2015px4}, ArduPilot \cite{ArduPilot}) for mission-critical tasks. For instance, AI-powered perception modules detect ArUco markers \cite{garrido2014automatic} for autonomous landing, while path-planning and obstacle avoidance modules use sensor data and map representations to compute feasible trajectories. These outputs are then interpreted by control nodes that interface with low-level flight controllers, executing smooth and responsive maneuvers. This hybrid design supports flexibility, scalability, and real-time coordination essential for deploying drones in dynamic environments.

\begin{figure}[!ht]
	\centering
	\includegraphics[width=0.5\linewidth]{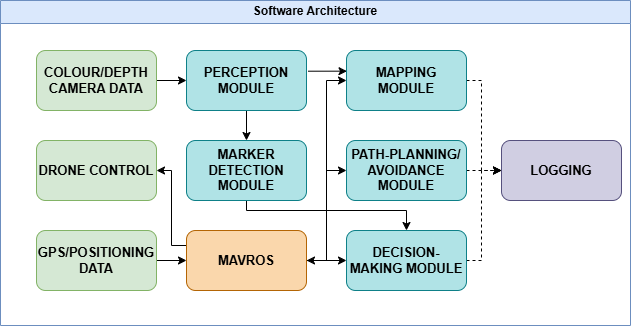}
        \caption{Components of a Popular Modular Autonomous Drone System \cite{schroder2025towards}.}
	\label{architecture} 
\end{figure}

Figure~\ref{architecture} illustrates a typical multi-module autonomous drone architecture. The system includes:
\begin{itemize}
\item \textbf{Perception Module}: Equipped with cameras, LiDAR, radar, GPS, IMUs, and ultrasonic sensors, drones continuously sense and interpret their environment. These modules process raw sensor data to detect obstacles, recognize landmarks or markers, and understand flight zones in real-time for navigation and landing.
\item \textbf{Localization and Mapping Module}: Using algorithms like SLAM (Simultaneous Localization and Mapping), GPS integration, and sensor fusion techniques, drones determine their precise position and map their surroundings, which is essential for safe and efficient navigation.
\item \textbf{Planning and Decision-Making Module}: These modules generate trajectories and make path-planning decisions based on mission objectives and environmental inputs. They include both global mission planners (e.g., waypoint generation \cite{causa2021multiple}) and local planners for real-time obstacle avoidance.
\item \textbf{Control Module}: Translating planned trajectories into actionable motor and actuator commands, control systems ensure stable and accurate flight, even in adverse weather or unforeseen situations.
\item \textbf{Communication Module}: These ensure seamless data exchange between onboard modules and ground stations or cloud-based services, enabling remote supervision, telemetry, and mission updates (e.g. aerial sensors equipped with Free Space Optics (FSO) \cite{kaushal2016optical} can transmit high volumes of image and video data to the command center through an onboard satellite communication subsystem).
\end{itemize}

This modular design paradigm, widely adopted in both academic and industrial systems, enhances the flexibility, scalability, and maintainability of autonomous drones. For example, frameworks such as ROS \cite{quigley2009ros} and PX4 \cite{meier2015px4} promote the development of decoupled modules that can be independently tested, updated, or replaced.

The widespread adoption of such modular systems in commercial and research contexts highlights the need for robust testing to ensure these complex components interact reliably and safely under diverse conditions. For example, industry-leading platforms such as Autoware \cite{Autoware} and Apollo \cite{Apollo} adopt similar modular architectures to manage the intricacies of autonomous driving and navigation. Their rigorous testing regimes underscore the critical role of validation processes in ensuring the performance and reliability of autonomous systems.

In this context, the necessity of a robust drone testing pipeline becomes clear. Testing must not only verify the correctness of individual modules but also validate system-wide behaviors and interactions. From handling corner cases in perception to ensuring low-latency control responses, comprehensive testing strategies are indispensable to guarantee safe deployment in real-world environments.

\subsection{Safety and Efficiency Imperatives}
As autonomous drones transition from research laboratories to real-world deployment, safety and efficiency have emerged as two of the most critical imperatives guiding their design, development, and testing. Whether delivering packages in urban environments, assisting in search-and-rescue missions, or conducting infrastructure inspections, autonomous drones are expected to operate reliably in complex, dynamic, and often unpredictable conditions. Any failure (e.g. a misclassified object, a navigation error, or a delayed response) could lead to catastrophic consequences, including property damage, mission failure, or endangerment of human lives \cite{shakhatreh2019unmanned}.

Unlike traditional software systems, where bugs may result in degraded user experience, failures in autonomous drones directly affect physical interactions with the environment. Thus, rigorous testing is required to: (1) mitigate risks, to identify and address potential failure points before real-world deployment \cite{fakhraian2023towards}; (2) ensure regulatory compliance, to meet aviation and safety standards, which are increasingly mandated by regulatory authorities \cite{EASA}; and (3) validate performance and efficiency, to guarantee that drones can fulfill mission requirements while optimizing resource usage, such as flight time, power consumption, and computation load \cite{annadata2025enhancing}.

The system under test (SUT) in this context comprises the core multi-module components of an autonomous drone system (as shown in Section \ref{multi-module}). These include the perception modules (e.g., sensor fusion, object detection), localization and mapping modules (e.g., SLAM, GPS-inertial integration), planning and decision-making modules (e.g., global mission planning, local obstacle avoidance), control modules (e.g., attitude and trajectory tracking), and communication modules (e.g., telemetry and remote command handling). Testing such a system is fundamentally more complex than traditional software testing. Not only must individual modules be verified for correctness, but their interactions, timing behavior, and real-time constraints must also be validated under realistic and often adversarial conditions. This inherent complexity necessitates a comprehensive and staged testing pipeline, supported by high-fidelity simulation environments, hardware-in-the-loop platforms, and structured real-world flight testing procedures \cite{sarkar2022framework}. Such infrastructure ensures that each module and their integration as a whole can be rigorously evaluated before deployment.

To comprehensively validate autonomous drones, testing must cover a broad spectrum of scenarios, ranging from routine operations to rare and dangerous edge cases. However, testing in the real world alone is impractical or unsafe for all situations. Simulation-based testing becomes essential to address these challenges, which offers several key advantages: (1) cost-effectiveness: scenarios that would be expensive or logistically complex to replicate physically (e.g., extreme weather, GPS loss) can be evaluated virtually \cite{Deng2023ADM}; (2) safety: dangerous situations, such as obstacle collisions or system failures, can be safely tested in simulated environments \cite{liang2025garl}; and (3) repeatability: simulated scenarios can be replayed with precision, enabling controlled experimentation and regression testing \cite{deng2022scenario}.

Given these requirements, a comprehensive testing pipeline has become the de facto approach in the autonomous drone industry. This pipeline typically includes the following four stages, each serving a distinct role in verifying system correctness and performance: software-in-the-loop (SIL) simulation testing, hardware-in-the-loop (HIL) testing, controlled real-world testing, and in-field testing. Through each of these stages, the autonomous drone undergoes increasingly realistic and challenging evaluations, ensuring that only well-tested systems proceed to mission-critical deployments.

To meet the stringent safety and reliability demands of autonomous drone systems, test-driven development (TDD) has emerged as a crucial engineering principle. Rather than retrofitting verification efforts after integration, TDD emphasizes continuous validation at each development stage. A representative example can be found in \cite{schroder2025towards}, where the authors adopt an iterative development and evaluation process that spans SIL, HIL, and real-world experiments. Through TDD, they successfully identify critical design flaws early such as visual detection instability or trajectory failures due to edge cases in dense environments, and resolve them with targeted improvements in marker detection, path planning, and onboard optimization.

Despite the benefits of TDD and staged validation, many commercial drone developers still rely heavily on real-world testing alone, often bypassing simulation, SIL, or HIL stages due to time pressure, resource constraints, or lack of infrastructure \cite{afzal2020study}. However, this practice can be both cost-inefficient and hazardous, as early-stage bugs may lead to catastrophic failures in uncontrolled environments. As highlighted in recent evaluations from both academia and industry, overlooking virtual testing stages not only undermines test coverage but also hampers scalability and reproducibility \cite{khatiri2023simulation}.

Against this backdrop, this paper plays an important role by systematically detailing the full autonomous drone testing pipeline from simulation to field validation, thereby promoting a more rigorous, staged, and test-driven development process for safer and more efficient UAV deployments. In the next section, we introduce an overview of the entire testing pipeline and represent how these stages fit together to form a systematic validation process for autonomous drone systems.

\subsection{Overview of Testing Pipeline}
\begin{figure}[!ht]
	\centering
	\includegraphics[width=0.7\linewidth]{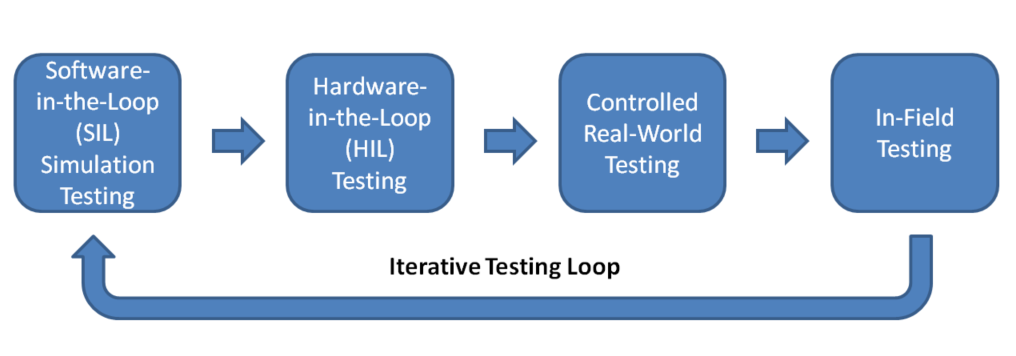}
        \caption{Overview of the autonomous drone testing pipeline.}
	\label{loop} 
\end{figure}

Developing a reliable and safe autonomous drone system requires much more than simply validating the functionality of individual modules. It demands a systematic and multi-stage testing pipeline that incrementally verifies system performance from isolated software modules to fully integrated real-world operations. Such a pipeline ensures that each level of complexity is thoroughly examined before advancing to the next stage, mitigating the risks of catastrophic failures in deployment. Figure~\ref{loop} illustrates the overall testing pipeline, which progresses through four major stages and highlights the iterative testing loop across the pipeline. This pipeline improves upon traditional distinctions by merging early-stage simulation and software validation into a unified category, and by clearly distinguishing between safe, controlled indoor testing and fully open-environment field trials.

\subsubsection{Software-in-the-Loop (SIL) Simulation Testing}
The first stage combines physics-based simulation (e.g., AirSim \cite{shah2018airsim}, CARLA \cite{dosovitskiy2017carla}, Gazebo \cite{Gazebo}, or SVL \cite{rong2020lgsvl}) with actual ROS \cite{quigley2009ros} modules or deep learning components to validate perception, localization, decision-making, and control logic in high-fidelity virtual environments. It allows repeatable and automated testing under varied scenarios with minimal safety risk and at scale, validating early prototypes and core algorithms. For example, vision-based autonomous driving systems (e.g., DeepDriving \cite{chen2015deepdriving}) use CNN-based perception modules that can be rigorously evaluated in simulation. Furthermore, simulation enables the reproduction of dangerous scenarios like collisions or sensor failures, and rapid prototyping and debugging without physical risk, allowing drones to be tested across a variety of weather conditions, lighting scenarios, and geographic landscapes. In addition to scenario creation, simulators can replay recorded sensor data and control commands from real-world test flights to validate new software releases. This approach is particularly valuable for regression testing, ensuring that updates or bug fixes do not introduce new issues.

\subsubsection{Hardware-in-the-Loop (HIL) Testing}
As the software modules mature, HIL testing integrates real flight controllers and onboard computing hardware components into the testing loop, while retaining a simulated physical environment. This stage introduces realistic communication delays, sensor noises, and actuator dynamics, providing a closer approximation to real-world operation. HIL testing is critical for measuring system latency to ensure real-time response capabilities, validating interfaces and interactions between software and hardware (e.g., flight controllers, GPS modules, motor drivers), and testing fault handling and recovery mechanisms under hardware-induced anomalies in a controlled environment \cite{VECTOR-HIL}. For example, Horri et al. \cite{horri2022tutorial} demonstrated that hardware-in-the-loop simulation, using Pixhawk 4 Mini autopilot \cite{Pixhawk} and MATLAB/Simulink \cite{Simulink} with virtual sensor models, is essential for validating robust UAV controllers prior to real-world flight.

\subsubsection{Controlled Real-World Testing}
This stage moves the drone into a physical but constrained environment such as an indoor warehouse, enclosed laboratory space or specialized flight arenas. These environments are typically equipped with safety measures including motion capture systems, safety nets, and tethering setups, that enable the safe evaluation of flight dynamics, perception modules, and control logic under real-world conditions. For instance, researchers at Harbin Institute of Technology constructed a 7m × 7m × 4m indoor drone testbed using a 16-camera motion capture system to validate multi-rotor flight control algorithms in a fully observable, repeatable environment \cite{NOKOV}. Similarly, commercial-grade safety net systems like those offered by Gorilla Netting allow for controlled drone operation in semi-outdoor or transitional spaces, helping mitigate the risk of unintended flyaways while still introducing environmental complexity such as wind or variable lighting \cite{GorillaNetting}. Controlled real-world testing strikes a balance between physical realism and operational safety, making it an indispensable stage for validating autonomous drone behavior before regulatory-sensitive field trials.

\subsubsection{In-Field Testing}
The final stage of the pipeline involves deploying the fully integrated drone into open, operational environments such as outdoor parks, industrial sites, or natural terrains, where wind, obstacles, and human presence introduce high variability and risk. This phase evaluates the system’s end-to-end performance, robustness, and ability to handle unforeseen circumstances. In this stage, adherence to local flight regulations, licensing, and safety protocols is critical. Real-world examples underscore the importance of this stage. Companies like Waymo \cite{Waymo} have reported extensive on-road testing for autonomous vehicle, covering over 1 million kilometers in 2020, with an average of 48,000 km between disengagements \cite{disengagement}. Similarly, Baidu's Apollo \cite{Apollo} has undertaken significant on-road and in-field testing initiatives before commercial rollout. According to the 2020 edition of the Beijing Autonomous Vehicles Road Test Report, Baidu achieved a cumulative mileage of over 2 million kilometers over a three-year period in Beijing, accounting for more than 90\% of the city's total testing mileage during that time \cite{BaiduRoadTest}.

By systematically progressing through these four stages, developers can ensure that autonomous drones meet the stringent safety, reliability, and performance requirements necessary for real-world deployment. This multi-layered approach forms the backbone of modern autonomous drone development pipelines and is essential for scaling such systems into critical, large-scale applications.

\subsection{Popular Autonomous Drone Components}
Autonomous drone systems rely on a range of core software components and infrastructure to support modular development, integration, and deployment. These frameworks abstract the complexities of drone control, sensor integration, and mission planning, while offering extensibility and interoperability. All features are essential for building and validating multi-module autonomous systems. Among these, three widely adopted and actively maintained components stand out: Robot Operating System (ROS) \cite{quigley2009ros}, PX4 Autopilot \cite{meier2015px4}, and ArduPilot \cite{ArduPilot}. Each plays a distinctive role in the testing pipeline and offers unique advantages and trade-offs. This section provides an overview of their roles, features, and differences to help practitioners select and configure components for their own autonomous drone projects.

\subsubsection{Robot Operating System (ROS)}
ROS \cite{quigley2009ros} is a flexible middleware framework that enables modular development of autonomous systems. In drone applications, ROS is particularly valued for its publish–subscribe communication model, its integration with simulation platforms (such as AirSim and Gazebo), and its compatibility with various autonomy algorithms. ROS allows developers to divide a complex autonomous drone system into independent functional nodes, each responsible for a specific task. This modularity is critical for debugging, testing, and scalability. Common ROS nodes used in an autonomous drone system include \cite{gustave2020functional}:
\begin{itemize}
\item \textbf{Marker Detector Node}: Processes camera input to detect visual markers such as ArUco tags. It publishes the marker’s pose (e.g., \texttt{geometry\_msgs/PoseStamped}) and serves as a core perception component in landing systems.
\item \textbf{Local Position Publisher Node}: Gathers and publishes real-time drone position data, often fusing GPS, IMU, or visual odometry inputs. This data is essential for navigation, path planning, and mission supervision.
\item \textbf{Waypoint Mission Node}: Manages the sequence of predefined waypoints that the drone must visit. It publishes waypoint targets and handles mission state transitions, enabling autonomous navigation through known routes.
\item \textbf{Navigation Node}: Performs path planning and obstacle avoidance based on a map or live sensor data. Algorithms such as A* \cite{hart1968formal}, RRT* \cite{karaman2011sampling}, or sampling-based motion planners may be used, and output is typically velocity or position setpoints.
\item \textbf{Drone Control Node}: Converts high-level navigation commands into low-level control inputs (e.g., thrust, yaw rate). It interacts with flight controllers like PX4 or ArduPilot via MAVROS \cite{MAVROS} or direct serial commands.
\end{itemize}

These nodes interact using standard ROS messages (e.g., \texttt{sensor\_msgs}, \texttt{nav\_msgs}, \texttt{geometry\_ msgs}) and services, enabling asynchronous and real-time communication. Developers can also use visualization tools such as RViz \cite{kam2015rviz} to inspect published topics (e.g., marker poses, planned paths) and use \texttt{rosbag} to record and replay test data.

One of the key advantages of using ROS for drone testing lies in its seamless integration with popular simulation tools such as AirSim \cite{shah2018airsim} and CARLA \cite{dosovitskiy2017carla}, which allows developers to prototype and evaluate autonomous behaviors in virtual environments. In addition, ROS facilitates both SIL and HIL testing through its standardized communication framework, including topics, services, and messages. This consistent messaging interface enables the modular testing of individual components as well as full-system integration. Furthermore, ROS benefits from a vast ecosystem and a highly active developer community, providing extensive libraries, tutorials, and third-party packages that accelerate development and troubleshooting.

Despite its flexibility, ROS presents several challenges and limitations in the context of autonomous drone testing. One major concern is its lack of native support for hard real-time control \cite{ROSrealtime}, which necessitates careful management of timing constraints, particularly in latency-sensitive applications such as flight control. Additionally, as drone systems scale in complexity with more modules, sensor inputs, and control loops, the architectural clarity of a ROS-based system can degrade \cite{shaukat2024trustworthy}. Without disciplined system design and modular structuring, it becomes increasingly difficult to maintain debuggability and ensure reliable performance across all interacting components.

Overall, ROS serves as the backbone for many academic and industrial autonomous drone projects due to its flexibility and ecosystem maturity. By structuring software in terms of well-isolated ROS nodes, developers can test, reuse, and replace components independently. This node-based architecture also facilitates the integration of advanced AI components such as neural networks for perception and neurosymbolic reasoning modules in future drone systems.

\subsubsection{PX4 Autopilot}
PX4 \cite{meier2015px4} is an open-source flight control software suite specifically designed for professional and research-grade UAV applications. It provides comprehensive support for multirotors, fixed-wing drones, vertical take-off and landing vehicles (VTOLs), and ground vehicles; autonomous mission planning; and real-time flight control and safety features. PX4 integrates seamlessly with ROS through the MAVROS \cite{MAVROS} interface, enabling advanced autonomy applications while handling low-level flight stabilization and control.

PX4 offers several advantages that make it well-suited for autonomous drone development and testing \cite{PX4UserGuide}. One of its key strengths is its built-in support for real-time flight control, enabled by hard real-time scheduling, which is essential for time-critical applications such as stabilization and navigation. Additionally, PX4 provides a comprehensive set of safety and recovery features, including geofencing, return-to-launch (RTL), and customizable failsafe modes, which enhance system robustness in unexpected conditions. PX4 is also highly compatible with widely used simulation environments such as Gazebo \cite{Gazebo} and AirSim \cite{shah2018airsim}, making it an effective platform for simulation-based testing and SIL testing during early development and integration stages.

While PX4 is a powerful and widely adopted flight control system, there are several considerations to keep in mind when integrating it into autonomous drone platforms. It is primarily optimized for aerial vehicles, meaning that developers working with ground-based systems or highly customized drone configurations may need to perform additional integration and adaptation work. Moreover, effective use of PX4 requires a solid understanding of its internal architecture, configuration parameters, and development workflows, which may present a learning curve for new users. Nevertheless, PX4 remains a popular choice for both research and commercial drone applications due to its advanced control capabilities and robust safety features, making it particularly well-suited for professional-grade systems.

\subsubsection{ArduPilot}
ArduPilot \cite{ArduPilot} is another powerful open-source autopilot platform supporting a wide variety of vehicles including multirotors, fixed-wing aircraft, rovers, boats, and submarines. Like PX4, ArduPilot offers extensive features for autonomous control, mission management, and failsafe operations.

ArduPilot offers several advantages that make it a compelling choice for a wide range of autonomous vehicle applications. One of its key strengths is its broad hardware support, which allows developers to deploy the system on diverse and non-standard platforms. In addition, ArduPilot features mature and robust flight control algorithms that have been extensively validated in both academic research and industrial deployments. This long-standing operational reliability makes it a trusted platform for mission-critical use cases. Furthermore, ArduPilot benefits from a well-established ecosystem, including comprehensive community-driven documentation and active developer forums \cite{ArduPilotCommunity}, which provide valuable support for new adopters and experienced users alike.

However, when considering ArduPilot for drone development, it is important to recognize certain limitations. One notable consideration is that ArduPilot does not offer as tight an integration with ROS as PX4, potentially requiring additional effort from developers when implementing advanced autonomy applications. Moreover, the configuration and tuning procedures within ArduPilot tend to be more manual and may necessitate a deeper level of technical expertise, especially for precise control and optimization tasks. These factors can contribute to a steeper learning curve and longer initial setup times compared to other platforms.

Notably, PX4 and ArduPilot are exclusive choices. While both provide excellent autopilot capabilities, developers must choose one based on their platform requirements and integration plans. Mixing them within the same drone platform is not feasible due to architectural differences. Table~\ref{Framework_Selection_Summary} summarizes key factors for selecting among the frameworks.

\begin{table*}
	\centering
	\caption{Framework Selection Summary}
	\label{Framework_Selection_Summary}
	\begin{threeparttable}
		\renewcommand\arraystretch{1.5}
		\newcolumntype{l}{>{\raggedright}p{0.15\textwidth}}
		\newcolumntype{A}{>{\raggedright}p{0.35\textwidth}}
		\newcolumntype{B}{>{\raggedright\arraybackslash}p{0.3\textwidth}}
		\newcolumntype{C}{>{\raggedright\arraybackslash}p{0.3\textwidth}}
		\newcolumntype{D}{>{\raggedright\arraybackslash}p{0.29\textwidth}}
		\scalebox{0.6}{
			\begin{tabular}{l A B C D}
				\toprule
				\textbf{Framework} & \textbf{Primary Role} & \textbf{Integration with ROS} & \textbf{Real-time Capability} & \textbf{Supported Testing Stages} \\ 
				\midrule
				ROS & Middleware for modular development and integration & Native & Soft real-time (with care) & Software-in-the-Loop, Hardware-in-the-Loop, Real-world \\
				\midrule
				PX4 & Real-time flight control and mission management & Excellent (via MAVROS) & Hard real-time & Software-in-the-Loop, Hardware-in-the-Loop, Real-world \\
				\midrule
				ArduPilot & Versatile autopilot for diverse vehicles & Moderate & Hard real-time & Software-in-the-Loop, Hardware-in-the-Loop, Real-world \\
				\bottomrule
		\end{tabular}}
	\end{threeparttable}
\end{table*}

In summary, selecting the appropriate framework or combination of frameworks is a critical first step when designing a robust drone testing pipeline. ROS serves as the integration backbone, while PX4 and ArduPilot provide low-level flight control and safety mechanisms. Understanding their roles, strengths, and limitations will directly impact the success of subsequent testing stages.

With these frameworks in place, the next step is to construct the actual pipeline and configure each testing stage for practical deployment. The following section provides a detailed, step-by-step guide on how to set up, run, and troubleshoot each stage of the testing pipeline, using the marker-based autonomous landing system as a running example.

\section{Step-by-Step Guide to Setting Up a Testing Pipeline}
As discussed in the previous section, a robust and well-structured testing pipeline is essential for validating autonomous drone systems at multiple levels of abstraction, ranging from virtual simulation models to fully integrated real-world deployments. This section provides a practical, step-by-step guide to setting up each core stage of the testing pipeline. It outlines how to initialize and configure the testing environment, addresses common issues and troubleshooting strategies, and presents real-world examples, with a particular focus on the marker-based autonomous landing system. By the end of this section, readers will gain both a conceptual framework and hands-on guidance for replicating and tailoring their own autonomous drone testing pipelines.

\subsection{Software-in-the-Loop (SIL) Simulation Testing}
Simulation-based testing serves as the foundational stage in the autonomous drone testing pipeline, offering a safe and flexible environment for early-stage validation. It enables developers to test algorithms and system behaviors without risking physical hardware or the surrounding environment. This approach is particularly valuable for exploring and validating rare or hazardous scenarios such as emergency landings or GPS signal loss that would be difficult or unsafe to replicate in real-world conditions. Additionally, simulation allows for rapid iteration and debugging without the logistical and financial burdens of repeated field trials. It also facilitates the generation of synthetic datasets, which can be used to train and evaluate perception modules under controlled and diverse conditions.

When selecting a simulation platform for autonomous drone development, two primary types of environments are typically considered: physics-based simulators such as AirSim \cite{shah2018airsim} and Gazebo \cite{Gazebo}. Among them, AirSim, developed by Microsoft Research and built on the Unreal Engine \cite{UnrealEngine}, is particularly well-suited for drone simulation due to its high-fidelity physics, photorealistic rendering, and strong compatibility with ROS. Gazebo also offers robust integration with ROS, making it a popular alternative. However, given AirSim’s visual realism and flexible API, this work focuses on simulation-based testing using the AirSim + ROS configuration.

To ensure smooth execution of high-fidelity simulation environments, the recommended hardware specifications are shown in Table~\ref{Recommended_System_Configuration}. It is important to note that insufficient RAM or GPU performance may lead to degraded simulation quality, including frame drops, input lag, and delays in sensor emulation, particularly during high-fidelity perception or control testing. For Linux-based development environments, NVIDIA GPUs are strongly recommended due to their robust driver support and compatibility with CUDA, which is essential for accelerated image processing and deep learning tasks commonly used in vision-based modules. Although macOS is supported by AirSim, it currently lacks full feature parity with Windows and Linux versions, and may exhibit limited functionality or reduced plugin support when integrating with ROS or Unreal Engine-based extensions.

\begin{table*}[!ht]
	\centering
	\caption{Recommended System Configuration for Simulation-based Testing}
	\label{Recommended_System_Configuration}
	\begin{threeparttable}
		\renewcommand\arraystretch{1.5}
		\newcolumntype{l}{>{\raggedright}p{0.3\textwidth}}
		\newcolumntype{A}{>{\raggedright}p{0.45\textwidth}}
		\newcolumntype{B}{>{\raggedright\arraybackslash}p{0.25\textwidth}}
		\newcolumntype{C}{>{\raggedright\arraybackslash}p{0.35\textwidth}}
		\newcolumntype{D}{>{\raggedright\arraybackslash}p{0.13\textwidth}}
		\scalebox{0.6}{
			\begin{tabular}{l A B C D}
				\toprule
				\textbf{Operating System} & \textbf{Processor} & \textbf{Memory (RAM)} & \textbf{Graphics Card} & \textbf{Disk Space} \\ 
				\midrule 
				Windows 10 (64-bit) & Quad-core Intel or AMD, $\geq$ 2.5 GHz & 8 GB & DirectX 11/12 compatible & $\geq$ 100 GB \\
				\midrule
				macOS Big Sur or later & Quad-core Intel, $\geq$ 2.5 GHz & 8 GB & Metal 1.2 compatible & $\geq$ 100 GB \\
				\midrule
				Ubuntu 18.04 LTS & Quad-core Intel or AMD, $\geq$ 2.5 GHz & 32 GB & NVIDIA GeForce GTX 960 & $\geq$ 100 GB \\
				\bottomrule
		\end{tabular}}
	\end{threeparttable}
\end{table*}

When deploying simulation-based testing within a virtual machine (VM) environment such as when using ROS and AirSim on a cloud-hosted system or local hypervisor, additional constraints must be taken into account. Unlike native installations, VM setups introduce additional performance overhead and often demand higher hardware specifications. In particular, it is recommended to allocate at least 16 GB of memory and a minimum of 150 GB of disk space to ensure smooth compilation and execution of the entire software stack, including AirSim, Unreal Engine assets, ROS workspaces, and sensor datasets. Moreover, if the VM is configured with a virtual GPU instead of direct access to a physical GPU, graphical performance will be significantly degraded. In such cases, the rendering overhead introduced by AirSim can overwhelm the system, causing lag or launch failures. To mitigate this, it is important to start the AirSim environment first, allowing the Unreal Engine instance to initialize fully before launching additional ROS nodes or visualization tools. This staging sequence reduces contention for virtualized resources and improves stability during testing.

\subsubsection{SIL Simulation Testing Setup}
The following sections detail the setup process and illustrate how to connect and operate the simulation environment for effective testing.

\paragraph{Install AirSim and Unreal Engine}
To begin using AirSim for high-fidelity simulation, developers are required to build the simulator from source.

\begin{enumerate}
\item \textit{Install Unreal Engine.} Download Unreal Engine source code (version 4.27 recommended for compatibility) and build it.
\begin{lstlisting}[language=bash]
 ./Setup.sh
 ./GenerateProjectFiles.sh
 make
\end{lstlisting}

\item \textit{Install AirSim.} After installing Unreal Engine, the next step is to obtain and build AirSim from its official GitHub repository. By default, AirSim builds using Clang 8, as this compiler version is known to be compatible with Unreal Engine 4.27. The included setup script will automatically handle the installation of the correct versions of CMake, LLVM/Clang, and Eigen on supported platforms. This ensures that the build environment is consistent and reproducible across development machines.
\begin{lstlisting}[language=bash]
 git clone https://github.com/Microsoft/AirSim.git
 cd AirSim
 ./setup.sh
 ./build.sh
\end{lstlisting}
\end{enumerate}

\paragraph{Connect AirSim with ROS using AirSim ROS Wrapper}
To enable perception, planning, and control modules to interact with the simulated drone, AirSim must be connected to a ROS environment. This is typically done using the \texttt{airsim\_ros\_pkgs} bridge, which exposes AirSim data (e.g., camera, IMU, GPS, control interfaces) through standard ROS topics and services.

Before proceeding, ensure that a compatible version of ROS is installed based on your operating system: For Ubuntu 16.04, install ROS Kinetic; For Ubuntu 18.04, install ROS Melodic; For Ubuntu 20.04, install ROS Noetic. This work focuses on ROS Noetic on Ubuntu 20.04, which is currently the most widely supported configuration for AirSim and ROS integration.

On Ubuntu 20.04, installing ROS Noetic involves the following commonly used steps:

\begin{enumerate}
\item First, make sure your package index is up-to-date.
\begin{lstlisting}[language=bash]
 sudo apt update
\end{lstlisting}

\item Add the ROS package repository.
\begin{lstlisting}[language=bash]
 sudo sh -c 'echo "deb http://packages.ros.org/ros/ubuntu $(lsb_release -sc) main" > /etc/apt/sources.list.d/ros-latest.list'
\end{lstlisting}

\item Set up the ROS key.
\begin{lstlisting}[language=bash]
 sudo apt install curl # if you haven't already installed curl
 curl -s https://raw.githubusercontent.com/ros/rosdistro/master/ros.asc | sudo apt-key add -
 sudo apt update
\end{lstlisting}

\item Install the full desktop version. For most development environments, it is recommended to install the full ROS Noetic desktop distribution, which includes essential GUI tools such as \texttt{rviz} \cite{kam2015rviz} and \texttt{rqt}. However, for onboard deployment or resource-constrained platforms (e.g., embedded systems or Jetson-based drones), installing \texttt{ros-noetic-ros-base} may be preferable, as it excludes GUI-related components and can significantly reduce both build time and system resource usage. Additionally, while many essential packages such as OMPL (Open Motion Planning Library) \cite{sucan2012open} are included by default in the desktop distribution, it is important to note that in virtual machine environments, the \texttt{ros-noetic-ompl} package may not be preinstalled. This can lead to build failures, particularly for planning-related modules. To avoid such issues, explicitly install OMPL using:
\begin{lstlisting}[language=bash]
 sudo apt install ros-noetic-desktop-full ros-noetic-ompl
\end{lstlisting}

\item Add ROS environment setup to your shell.
\begin{lstlisting}[language=bash]
 sudo chmod +x /opt/ros/noetic/setup.bash
 echo "source /opt/ros/noetic/setup.bash" >> ~/.bashrc
 source ~/.bashrc
\end{lstlisting}

\item Initialize \texttt{rosdep} and update dependencies. \texttt{rosdep} is used to install supporting system packages automatically.
\begin{lstlisting}[language=bash]
 sudo apt install python3-rosdep
 sudo rosdep init
 rosdep update
\end{lstlisting}

\item Install toolchain and dependencies for building packages. To create and manage ROS work-spaces effectively, you need several development tools that are distributed as separate packages. For example, \texttt{rosinstall} is a commonly used command-line utility that allows you to download multiple ROS package source trees in a single step. \texttt{wstool} is used to manage and update the source directories within a workspace. For building packages, \texttt{catkin\_tools} is recommended, as it provides a more efficient and user-friendly alternative to the traditional \texttt{catkin\_make}.
\begin{lstlisting}[language=bash]
 sudo apt install python3-rosinstall python3-rosinstall-generator python3-wstool build-essential
 pip install "git+https://github.com/catkin/catkin_tools.git#egg=catkin_tools"
\end{lstlisting}

\item Build ROS package. Before building ROS packages, ensure that your default GCC version is 8 or higher, as earlier versions may lead to compilation errors.
\begin{lstlisting}[language=bash]
 gcc --version
 sudo apt-get install gcc-8 g++-8 # install GCC 8 if your default version is not 8 or above
 cd ros
 catkin build
\end{lstlisting}

\item Once ROS Noetic is installed, launch the ROS node and connect to the AirSim simulation environment. Verify basic telemetry (pose, images, IMU data) is published in ROS topics.
\begin{lstlisting}[language=bash]
 roslaunch airsim_ros_pkgs airsim_node.launch
 rostopic list
\end{lstlisting}
\end{enumerate}

\paragraph{Setting Up a ROS Workspace and Package}
A ROS workspace is a directory where you can build, modify, and organize multiple ROS packages. It typically contains source code, build files, and environment setup scripts. The default workspace layout in ROS (using \texttt{catkin}) includes the following:
\begin{lstlisting}
 catkin_ws/
 |-- src/       <- your source packages go here
 |-- build/     <- auto-generated build files (after compilation)
 |-- devel/     <- development space with environment setup scripts
\end{lstlisting}

Different from \texttt{airsim\_ros\_pkgs}, which serves as the official ROS interface package for AirSim, a custom ROS package allows you to build application-specific logic and experimental modules on top of it. While \texttt{airsim\_ros\_pkgs} primarily provides standardized ROS topics and services for accessing sensor data and sending control commands within the AirSim environment, your custom package introduces mission-specific functionality through custom nodes, launch files, and configurations. This custom package depends on \texttt{airsim\_ros\_pkgs} but extends its capabilities to support tailored drone behaviors. An overview comparison is presented in Table~\ref{custom_package}.

\begin{table*}[!ht]
	\centering
	\caption{Overview Comparison Between \texttt{airsim\_ros\_pkgs} and a Custom Package}
	\label{custom_package}
	\begin{threeparttable}
		\renewcommand\arraystretch{1.5}
		\newcolumntype{A}{>{\raggedright}p{0.3\textwidth}}
		\newcolumntype{B}{>{\raggedright\arraybackslash}p{0.45\textwidth}}
		\newcolumntype{C}{>{\raggedright\arraybackslash}p{0.46\textwidth}}
		\scalebox{0.6}{
			\begin{tabular}{A B@{\hskip 0.6in}C}
				\toprule
				\textbf{Feature} & \textbf{\texttt{airsim\_ros\_pkgs}} & \textbf{Custom Package} \\ 
				\midrule 
				\textbf{Type} & Official ROS wrapper & User-defined ROS package \\
				\midrule
				\textbf{Purpose} & Bridge between AirSim and ROS & Encapsulate specific application logic for your drone project \\
				\midrule
				\textbf{Maintained By} & Microsoft/AirSim team and contributors & You (or your team) \\
                \midrule
				\textbf{Key Functionality} & Publishes sensor data, handles control APIs, syncs simulation with ROS & Implements custom nodes (e.g., perception, planning, control) \\
                \midrule
				\textbf{Topics Provided} & \texttt{/airsim\_node/*} topics: IMU, camera, lidar, GPS, etc. & Custom topics like \texttt{/drone\_1/marker\_pose}, \texttt{/landing\_cmd} \\
                \midrule
				\textbf{Launch Files} & \texttt{airsim\_node.launch} & \texttt{demo\_sitl.launch}, \texttt{drone\_simulation\_with\_args.launch} \\
                \midrule
				\textbf{Dependencies} & Depends on AirSim + AirLib + ROS core packages & Depends on \texttt{airsim\_ros\_pkgs}, plus your control/perception stack \\
                \midrule
				\textbf{Extensibility} & Low–focus is on bridging AirSim only & High—you define functionality, parameters, behaviors \\
				\bottomrule
		\end{tabular}}
	\end{threeparttable}
\end{table*}

To implement and test autonomous drone behaviors in simulation, a well-structured ROS package is essential. In this work, we introduce a custom package named \texttt{drone\_simulation} as an example, which encapsulates mission logic, perception, and control modules tailored for integration with AirSim. The overall package layout and key directories are briefly outlined here, with full details provided in Appendix~\ref{appendix:pkg-layout}.

A ROS launch file is an XML-based configuration script used to automate the startup of multiple ROS nodes, load parameters, and manage runtime options. In \texttt{drone\_simulation}, launch files play a critical role in combining perception and control modules for test orchestration. We first demonstrate a basic single-node launch configuration (Appendix~\ref{appendix:marker-launch}), and then extend it to a multi-module setup (Appendix~\ref{appendix:simulation-launch}) to execute a complete mission scenario from marker detection to autonomous landing.

In ROS, a node is often defined within a Python script, typically located in the \texttt{scripts/} directory of a ROS package. These scripts are responsible for implementing core functionalities such as perception, control, data processing, or service interaction. For instance, the \texttt{marker\_detector.py} script (Appendix~\ref{appendix:marker-detector}) serves as a ROS node that subscribes to a camera image topic, detects ArUco markers, estimates their 3D pose, and publishes this information as a \texttt{PoseStamped} message. Python-based ROS nodes like this are executed using \texttt{rosrun} or launched through \texttt{.launch} files. To function properly, they must be granted executable permissions and have all required dependencies declared in the \texttt{package.xml} file (Appendix~\ref{appendix:package-deps}). When integrated with ROS launch files and modular system configurations, Python scripts provide a highly flexible foundation for rapid prototyping and development in autonomous drone systems. In performance-critical scenarios, these Python nodes are often re-implemented in C++ (as \texttt{.cpp} files) to achieve greater execution speed and system efficiency.

The following steps outline how to set up a ROS workspace and a custom package:

\begin{enumerate}
\item Create the workspace. \texttt{catkin\_ws} is the name of your workspace (you can name it as you like). \texttt{catkin build} builds all packages in the \texttt{src} folder (currently no packages) and generates the \texttt{build/} and \texttt{devel/} folders.
\begin{lstlisting}[language=bash]
 mkdir -p ~/catkin_ws/src
 cd ~/catkin_ws
 catkin build
\end{lstlisting}

\item Add workspace to environment variables. Append the workspace setup to your \texttt{.bashrc} so ROS can recognize your workspace automatically. This ensures that every new terminal session loads the workspace environment.
\begin{lstlisting}[language=bash]
 echo "source ~/catkin_ws/devel/setup.bash" >> ~/.bashrc
 source ~/.bashrc
\end{lstlisting}

\item Create a custom package in a ROS context. Navigate into the \texttt{src} folder and create a new package. \texttt{drone\_simulation} is your package name (replace with your own). The remaining arguments are dependencies required by your package (you can add more later).
\begin{lstlisting}[language=bash]
 cd ~/catkin_ws/src
 catkin create pkg drone_simulation rospy std_msgs geometry_msgs sensor_msgs
\end{lstlisting}

\item Add your code and resources. Inside your new package, \texttt{launch/} has launch files to start nodes and configurations, \texttt{scripts/} has Python scripts or ROS nodes. Then give execution permission to all Python ROS nodes.
\begin{lstlisting}[language=bash]
 cd ~/catkin_ws/src/drone_simulation
 mkdir launch scripts
 chmod +x scripts/*.py
\end{lstlisting}

\item Build the workspace. Go back to the root of your workspace and compile. After building, refresh the workspace environment.
\begin{lstlisting}[language=bash]
 cd ~/catkin_ws
 catkin build
 source devel/setup.bash
\end{lstlisting}

\item In practice, the integration begins by launching \texttt{airsim\_ros\_pkgs}, which establishes the communication bridge between AirSim and ROS topics.
\begin{lstlisting}[language=bash]
 roslaunch airsim_ros_pkgs airsim_node.launch
\end{lstlisting}
Then you run your custom package to start your control/perception/logic stack that subscribes to AirSim data and publishes control commands.
\begin{lstlisting}[language=bash]
 roslaunch drone_simulation demo_sitl.launch
\end{lstlisting}
\end{enumerate}

\subsubsection{Case Study: SIL Simulation Testing for Marker-Based Autonomous Landing System}
\begin{figure}[!ht]
	\centering
	\includegraphics[width=0.8\linewidth]{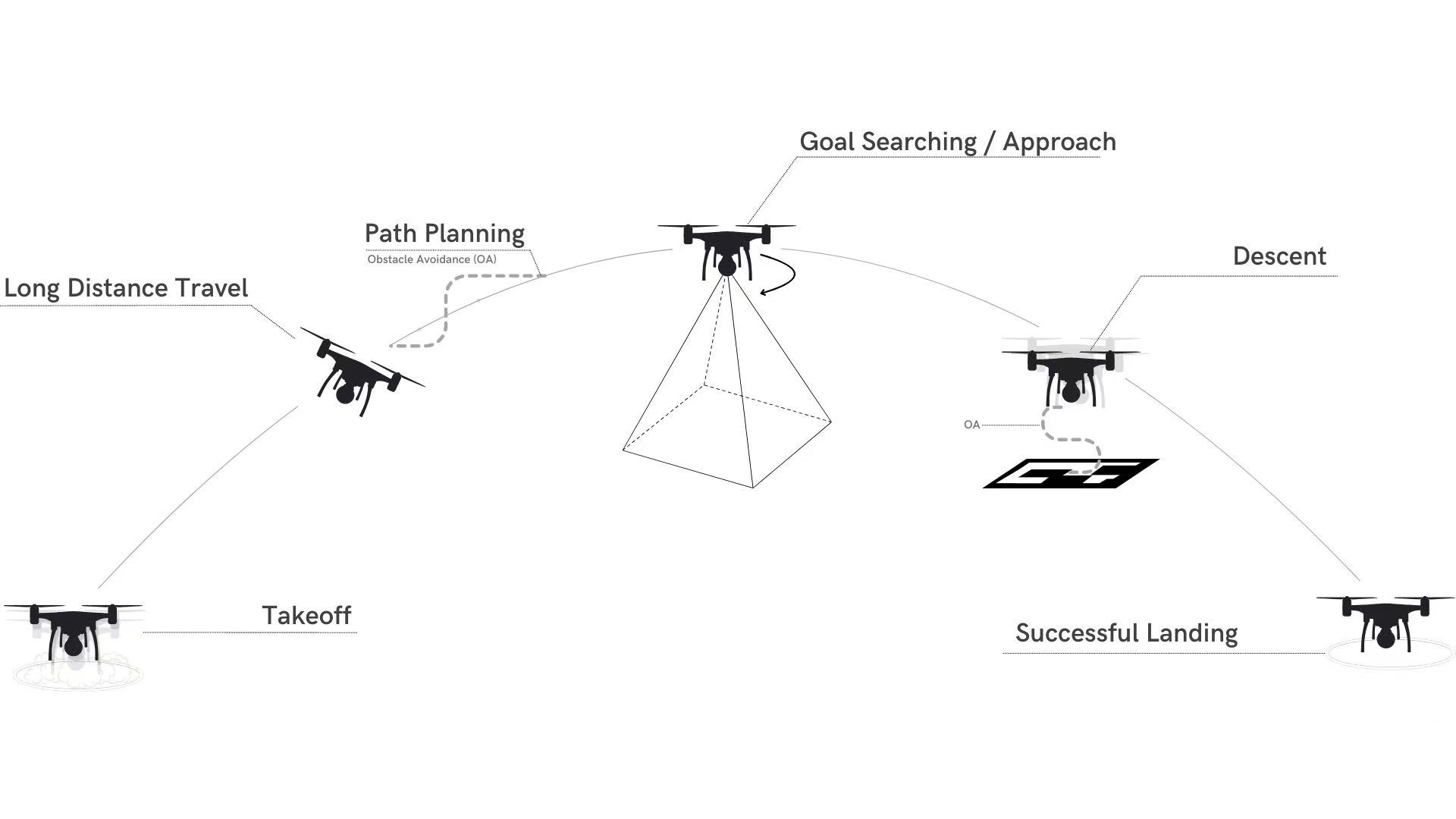}
        \caption{Overview of the marker-based autonomous landing process.}
	\label{landing} 
\end{figure}

Software-in-the-Loop (SIL) simulation testing plays a crucial role in the early-stage validation of autonomous drone systems, particularly when testing complex subsystems like marker-based landing logic. In this case study, we examine how SIL testing was used to assess the robustness and evolution of a marker-based autonomous landing system, which integrates perception, mapping, and planning modules within a ROS-based architecture. The overall process of the marker-based autonomous landing system is illustrated in Figure~\ref{landing}, which visualizes the sequential flight phases from takeoff through long-distance travel, marker detection, and descent, culminating in a successful landing on the visual target.

\paragraph{SIL Testing Setup and Experimental Design}
The system under test is a modular landing framework that employs ArUco markers \cite{garrido2014automatic} for visual guidance during final descent. The core components include a marker detection module, a mapping module, and a trajectory planning module. The testing was conducted in the AirSim simulator, integrated with ROS and PX4. To evaluate system performance and iterate design improvements, three successive generations of the landing system, MLS-V1, MLS-V2, and MLS-V3 were tested under identical simulation scenarios. These scenarios varied in lighting conditions, obstacle placement, and terrain textures, all rendered using high-fidelity Unreal Engine \cite{UnrealEngine} environments.

\paragraph{Results and Iterative Improvements}
Quantitative results from the SIL tests showed meaningful progression across the three generations:

\begin{enumerate}
\item \textit{MLS-V1:} The first generation system based on OpenCV \cite{bradski2000opencv} had the highest failure rate due to frequent loss of marker detection during descent and poor trajectory planning in cluttered environments.
\item \textit{MLS-V2:} The second generation system using TPH-YOLOv5 \cite{zhu2021tph} for marker detection and EGO-Planner \cite{zhou2020ego} for path planning introduced improved trajectory logic that reduced collisions, though it still exhibited occasional failures during landing under variable lighting.
\item \textit{MLS-V3:} The third generation system adopting OctoMap \cite{hornung2013octomap} mapping and RRT* \cite{karaman2011sampling} path planning algorithm achieved the best overall performance, with a significant reduction in both collision rate and failed landings, indicating successful iteration over previous designs.
\end{enumerate}

\begin{figure}[!ht]
    \centering
    \subfloat[OpenCV failed marker detection.]{\includegraphics[width=5cm]{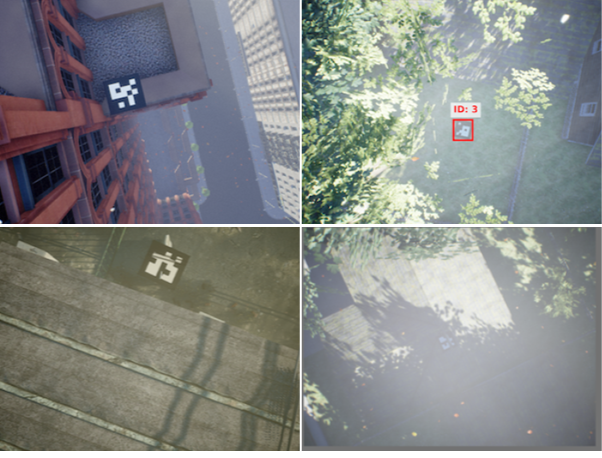}
    \label{fig:opencv}}
    \hspace{1cm}
    \subfloat[A* planner struggles large obstacle.]{\includegraphics[width=5cm]{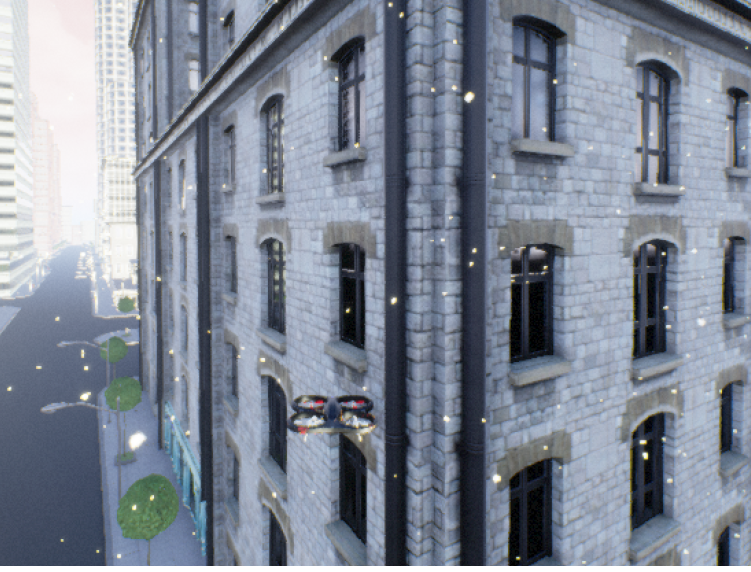}
    \label{fig:collision}}
    \captionsetup{font=small}
    \caption{Illustrative failures observed during SIL simulation testing of a marker-based autonomous landing system \cite{schroder2025towards}.}
    \label{fig:failures}
\end{figure}

\paragraph{Limitations}
Several critical issues were uncovered during SIL simulation testing:

\begin{enumerate}
\item \textit{Marker Detection Module:} OpenCV-based ArUco detection failed under high-altitude views, partial occlusion (e.g., tree branches or shadows), and lighting conditions involving glare or overexposure. This led to misidentification or complete loss of visual landmarks mid-descent. As shown in Figure~\ref{fig:opencv}, the OpenCV-based marker detection module failed to identify the ArUco marker at high altitude due to partial occlusion and harsh lighting conditions (e.g., sun glare).

\item \textit{Mapping Module:} There was a trade-off between occupancy grid resolution and memory usage. High-resolution maps enabled better path planning but quickly exhausted system memory during long missions.

\item \textit{Path Planning Module:} The A* \cite{hart1968formal} planner struggled with large or dynamic obstacles. In some cases, the UAV attempted to fly through temporarily undetected objects (e.g., foliage), which resulted in entrapment or failure to reach the landing zone. In Figure~\ref{fig:collision}, the A* path planner struggled to generate a valid trajectory around a large obstacle, causing collision during turning action.
\end{enumerate}

\paragraph{Simulator Capabilities and Insights}
The use of AirSim as the simulation backend was instrumental in discovering these issues. Compared to traditional platforms like Gazebo, AirSim offered superior visual fidelity, enabling realistic reproduction of sensor behavior under various environmental conditions. As a plugin for Unreal Engine, AirSim also allowed the development team to construct diverse test environments ranging from rural farmlands to urban rooftops by simply reusing assets from Unreal game projects. Moreover, AirSim supported ROS integration for real-time simulation of sensor streams and control commands, thereby enabling closed-loop testing of the full landing pipeline.

\paragraph{Key Challenges in SIL Simulation Testing}
Software-in-the-Loop (SIL) simulation remains a foundational tool in autonomous drone system validation. However, its effectiveness is increasingly constrained by the fundamental divide between data-driven components and rule-based logic within typical drone architectures. On one hand, statistical and machine learning-based modules, particularly those relying on deep neural networks, provide strong capabilities for perception and decision-making in high-dimensional environments. However, these models are often treated as black boxes, making it difficult to ensure explainability or formally verify correctness \cite{csahin2025unlocking}. While blackbox testing methods such as random input generation or adversarial sampling can reveal failure modes, they cannot guarantee full behavioral correctness. Traditional formal verification techniques like theorem proving, model checking, and static analysis fail to scale to these non-symbolic components due to their lack of logical structure and semantics \cite{badings2023probabilities}.

Conversely, rule-based symbolic components (e.g., mission planners or control logic encoded in PX4/ArduPilot) offer high transparency and logical rigor but lack adaptability in continuous, uncertain environments \cite{jacobson2025integrating}. This rigidity makes them brittle in scenarios involving partial observations, dynamic obstacles, or noisy sensor inputs. As a result, neither paradigm in isolation can comprehensively address the safety and robustness requirements of autonomous drone operations, especially in edge-case scenarios that lie outside typical operating envelopes.

Beyond this architectural split, SIL simulation testing faces inherent challenges stemming from limitations in current simulation environments. For instance, while simulators such as AirSim and Gazebo enable cost-effective, iterative testing, they often fall short in modeling multi-agent interactions \cite{suo2021trafficsim}. In our own testing, the marker detection module struggled to detect and react to dynamic obstacles such as pedestrians or vehicles, which are critical scenarios for urban deployment but poorly modeled due to computational constraints and simplified agent behaviors.

Additionally, these simulation tools typically rely on generalized models rather than detailed, system-specific characterizations. For example, in our AirSim-based SIL pipeline, perception modules processed camera images and GPS signals but lacked access to high-resolution LiDAR point clouds necessary for detailed 3D spatial reasoning. This compromises the ability to simulate tasks like precision landing or obstacle-aware navigation in cluttered environments. Such limitations reduce predictive accuracy and restrict the transferability of insights from simulation to field deployment.

In summary, while SIL simulation remains an essential step in the autonomous drone testing pipeline, its utility is limited by both systemic architectural challenges and technological constraints in current simulators. By embracing neurosymbolic methods, co-simulation architectures, and digital twin technologies, future SIL frameworks will be better equipped to support safety-critical validation and mission-specific reliability assessment.

\subsection{Hardware-in-the-Loop (HIL) Testing}
Hardware-in-the-Loop (HIL) testing represents a critical advancement over Software-in-the-Loop (SIL) testing by incorporating actual hardware components into the simulation environment, thereby bridging the gap between purely software-based simulations and real-world operation. While SIL testing effectively validates software logic and algorithm correctness in isolation, it inherently overlooks the complexities, imperfections, and unpredictable behaviors associated with physical hardware. HIL testing addresses this limitation by integrating real hardware components such as flight controllers, onboard sensors, and actuators into the testing loop. This integration enables comprehensive validation of hardware-software interfaces, identification of system latency and real-time performance constraints, and evaluation of potential hardware-related failure modes, including sensor noise and actuator delays. Additionally, HIL testing provides insights into critical operational aspects such as power consumption profiles, communication latencies, and precise timing requirements. For autonomous drone systems, which depend heavily on rapid and accurate interaction between software algorithms and physical hardware, HIL testing is indispensable. It ensures system-level reliability and robustness under realistic conditions, significantly reducing risk and enhancing overall safety prior to field deployment.

A typical Hardware-in-the-Loop (HIL) testing setup for autonomous drones involves integrating physical drone hardware components with simulation and computational resources. Key hardware components typically include a flight controller (commonly PX4 \cite{meier2015px4} or ArduPilot \cite{ArduPilot}) responsible for executing low-level control commands, stabilizing the drone, and interfacing directly with the higher-level software stack. Sensors, either as actual physical devices (e.g., inertial measurement units (IMU), GPS modules, cameras) or as carefully designed emulators, provide critical real-time input for validation purposes. The testing environment also includes a host computer that runs the software stack, typically based on ROS. This software component executes essential perception, planning, and mission logic modules. Real-time communication between the host computer and the drone’s hardware components is facilitated through interfaces such as Serial, MAVLink, or Ethernet protocols. Furthermore, a simulation environment or dataset playback mechanism is integrated into the setup, emulating various external world scenarios and ensuring realistic conditions for the drone's response validation. By closely replicating real operational dynamics, this comprehensive HIL arrangement ensures rigorous and realistic validation of hardware-software interactions, thus significantly enhancing system reliability prior to real-world deployment.

\subsubsection{HIL Testing Setup}
To illustrate the practical implementation of Hardware-in-the-Loop (HIL) testing in autonomous drone systems, we present a representative setup based on the PX4 autopilot firmware integrated with ROS. PX4 is a widely adopted open-source flight control stack, offering high configurability, robust performance, and strong support for HIL simulation. In this configuration, PX4 runs on a real flight controller connected to a host PC, where ROS handles higher-level tasks such as perception, planning, and mission execution. The two systems communicate over the MAVLink protocol, enabling real-time exchange of sensor data, actuator commands, and system state information. By combining PX4’s low-level flight control capabilities with ROS’s modular software architecture, this HIL setup provides a reliable and flexible testing environment that closely approximates the full operational behavior of the drone before transitioning to physical flight trials.

\paragraph{Prepare Flight Controller Hardware}
The first step in setting up a Hardware-in-the-Loop (HIL) testing environment involves configuring the flight controller hardware that will execute the drone's low-level control logic. A commonly used option is a development board running the PX4 firmware, such as the Pixhawk \cite{Pixhawk} or NVIDIA Jetson Nano \cite{JetsonNano} with PX4 integration. These controllers provide real-time flight control execution and support standard interfaces for sensor and actuator integration. The flight controller is physically connected to the host computer where the high-level autonomy stack runs via USB for serial communication or through a telemetry radio module that supports MAVLink protocol. This connection establishes the real-time communication channel necessary for transmitting sensor data, receiving actuator commands, and synchronizing system states. Proper power supply and isolation precautions must also be taken during hardware setup to prevent unintended activation or hardware damage during early-stage testing.

\paragraph{Connect Software Stack (ROS) to Flight Controller}
Once the flight controller is physically connected and powered, the next step is to establish software-level communication between the ROS-based autonomy stack and the flight controller. This is accomplished using the MAVROS package \cite{MAVROS}, an official ROS interface for MAVLink-enabled autopilot systems such as PX4. MAVROS acts as a bridge, translating ROS messages into MAVLink commands and vice versa, enabling seamless interaction between ROS nodes and the flight controller.

To connect ROS nodes to the flight controller, follow these steps:

\begin{enumerate}
\item \textit{Install PX4.} Before establishing communication between ROS and the flight controller, it is essential to ensure that the PX4 autopilot firmware is correctly installed and configured on the hardware. When working with ROS Noetic on Ubuntu 20.04, first clone the PX4-Autopilot repository:
\begin{lstlisting}[language=bash]
 git clone https://github.com/PX4/PX4-Autopilot.git --recursive
\end{lstlisting}
Then run the \texttt{ubuntu.sh} to build the firmware:
\begin{lstlisting}[language=bash]
 bash ./PX4-Autopilot/Tools/setup/ubuntu.sh --no-sim-tools --no-nuttx
\end{lstlisting}
Install the following additional dependencies:
\begin{lstlisting}[language=bash]
 sudo apt-get install protobuf-compiler libeigen3-dev libopencv-dev -y
\end{lstlisting}
Once compiled, the firmware is uploaded to the flight controller via USB.

\item \textit{Install MAVROS.} After flashing the PX4 firmware, the next step is to connect the ROS-based autonomy stack to the flight controller. This requires installing the MAVROS package on the host system, which serves as a communication bridge between ROS nodes and the PX4 system using the MAVLink protocol. MAVROS translates ROS messages into MAVLink commands and allows bidirectional communication of telemetry, control commands, and system states.
\begin{lstlisting}[language=bash]
 sudo apt-get install ros-noetic-mavros ros-noetic-mavros-extras ros-noetic-mavros-msgs
\end{lstlisting}
Then install the required geographic libraries:
\begin{lstlisting}[language=bash]
 wget https://raw.githubusercontent.com/mavlink/mavros/master/mavros/scripts/install_geographiclib_datasets.sh
 sudo bash ./install_geographiclib_datasets.sh
\end{lstlisting}

\item \textit{Initiate communication.} Once PX4 and MAVROS are installed, a typical launch configuration file such as \texttt{px4.launch} can be used to initiate communication. This launch file initializes the MAVROS node and configures parameters for serial connection, system ID, and frame settings. To verify that the connection is established successfully, developers can inspect the \texttt{/mavros/state} topic in ROS, which continuously publishes the flight controller’s current status including armed state, flight mode, and connection status:
\begin{lstlisting}[language=bash]
 roslaunch mavros px4.launch
 rostopic echo /mavros/state
\end{lstlisting}
\end{enumerate}

\paragraph{Simulate Sensor Inputs}
With the communication bridge established between ROS and the PX4-equipped flight controller, the next crucial step in the HIL testing process is to provide simulated sensor inputs that emulate the real-world environment. This is typically achieved using PX4’s Software-in-the-Loop simulation mode, which allows synthetic sensor data such as GPS, IMU, barometer, magnetometer, and visual odometry to be generated and streamed to the flight controller in real time.

In an HIL setup, SIL simulation can be configured to operate in conjunction with a high-fidelity simulation environment such as AirSim, Gazebo, or custom replay datasets. These environments simulate the drone's motion and physical interactions, producing realistic sensor outputs that are published to PX4 via MAVLink or direct sensor interface emulators. For example, GPS location updates and inertial measurements generated by the simulator are fed into the flight controller to drive control decisions, while visual input (e.g., camera streams) is simultaneously processed by perception nodes running in ROS.

The goal of this stage is to validate how the integrated system that comprising real flight hardware and software algorithms responds to dynamic, sensor-rich scenarios. It also facilitates testing of failure modes such as noisy measurements, sensor dropouts, or latency-induced errors. By simulating sensor inputs through SIL simulation in a controlled and repeatable manner, developers gain critical insights into the behavior and robustness of the drone’s perception and control pipelines before advancing to uncontrolled real-world testing environments.

\paragraph{Integrate Planning and Control Modules}
Once sensor inputs are simulated and the communication bridge between ROS and PX4 is functioning, the next step involves integrating the high-level planning and control modules. In a modular ROS-based architecture, these components are typically implemented as dedicated ROS nodes that calculate motion plans and generate control commands based on real-time sensor data and mission objectives. The planning module computes desired trajectories or waypoints, while the control module translates these into specific velocity or position setpoints.

These setpoints are published to MAVROS topics, such as \texttt{/mavros/setpoint\_position/local} or \texttt{/mavros/setpoint\_velocity/cmd\_vel\_unstamped}, depending on the mode of operation. MAV-ROS then relays the commands to the PX4 flight controller via MAVLink. PX4 processes these inputs within its control loops and generates corresponding actuator signals to simulate or drive the drone’s motors, gimbals, or servos.

This integration allows developers to validate the entire perception–planning–control pipeline in a realistic feedback loop. Errors such as trajectory instability, overshooting, or poor convergence can be detected early in HIL testing, enabling iterative refinement of planning logic and controller parameters. Moreover, performance metrics like tracking accuracy, system latency, and recovery behavior in edge cases can be quantitatively evaluated. This step is essential for ensuring that control outputs remain smooth, safe, and reliable before transitioning to controlled or in-field flight testing.

\paragraph{Monitor and Measure System Performance}
The final step in HIL testing focuses on monitoring and evaluating the performance of the integrated drone system under simulated operational conditions. This involves systematically recording and analyzing key runtime metrics that reflect the system’s responsiveness, stability, and robustness. Using ROS diagnostic tools (such as \texttt{rqt\_plot}, \texttt{rosbag}, or \texttt{rqt\_graph}) in combination with PX4’s onboard logging mechanisms (e.g., \texttt{ulog} files), developers can collect time-synchronized data across multiple subsystems for post-analysis.

A critical performance indicator is command latency, which is the time elapsed between the publication of control setpoints in ROS and the corresponding actuator response on the flight controller. High latency may indicate communication bottlenecks or processing inefficiencies that can compromise real-time behavior. Another important metric is the stability and precision of control execution, typically assessed by analyzing tracking performance and trajectory deviations over time. Sudden oscillations or overshooting may point to tuning issues within the control loop.

Additionally, HIL testing provides a safe environment to simulate and monitor system responses to fault conditions, such as sensor dropouts, communication loss, or unexpected actuator behavior. The ability to observe how the system detects, handles, and recovers from such failures is vital for verifying fault-tolerance mechanisms. By capturing and analyzing these metrics during HIL testing, developers gain a comprehensive understanding of system-level performance and can confidently refine the design before advancing to controlled or in-field deployments.

\subsubsection{Case Study: HIL Testing for Marker-Based Autonomous Landing System}
Following the extensive Software-in-the-Loop (SIL) simulations across multiple iterations of the marker-based landing system (MLS-V1 to MLS-V3), the development team transitioned to Hardware-in-the-Loop (HIL) testing to validate how well the software performed when interfacing with real hardware components. This transition aimed to assess system behavior under realistic sensor latencies, hardware constraints, and execution delays that are not observable in pure simulation environments.

\paragraph{HIL Testing Setup and Experimental Design}
The HIL testing environment integrated a CUAV X7+ Pro Pixhawk flight controller \cite{CUAV} running PX4 autopilot firmware, real-time perception and planning modules hosted on a NVIDIA Jetson Nano \cite{JetsonNano} 4GB companion computer with PX4 integration, and camera input fed from emulated onboard sensors. The Jetson Nano was set to operate in MAXN power mode, which allows the GPU and CPU to run at full power, ensuring maximum inference throughput. The ROS middleware coordinated communication between components via MAVROS \cite{MAVROS}. To meet real-time requirements, the trained vision model was optimized and deployed using TensorRT \cite{TensorRT}, NVIDIA’s high-performance deep learning inference runtime. Converting the original model into the TensorRT format allowed the system to achieve significantly lower latency and higher frame rates compared to its PyTorch or ONNX counterparts. This optimization was critical for ensuring that the system could consistently detect landing markers and generate control commands at the required 30 FPS. Controlled input stimuli such as camera images of ArUco markers and simulated IMU/GPS signals were injected into the system to emulate realistic flight scenarios, while actuator outputs were monitored to verify expected drone responses. In alignment with lessons from SIL testing, the experiments were designed to stress-test landing behavior during dynamic descent, under partial marker occlusion, GPS drift, and sensor noise. Testing was repeated across multiple missions to ensure reproducibility and observe edge-case failures.

\paragraph{Results and Iterative Improvements}
The integration of TensorRT inference on Jetson Nano in MAXN mode yielded practical benefits. It reduced latency in perception pipelines, resulting in faster pose estimation cycles. Moreover, it improved responsiveness in control loops due to consistent and deterministic marker detection. This setup enabled precise trajectory control during descent, with landing accuracy improving to within ±5 cm under standard conditions. Furthermore, the perception system remained reliable even under moderate motion blur and illumination changes, conditions in which less optimized inference pipelines would drop frames or misclassify marker IDs.

\paragraph{Limitations}
Despite performance gains, several issues emerged. First, heat generation from running MAXN mode required careful thermal management. Additionally, initial TensorRT conversion resulted in minor accuracy drop due to layer approximations, necessitating post-quantization fine-tuning. These issues emphasized the trade-off between maximizing throughput and maintaining deterministic timing, especially in tightly coupled perception-control loops.

\paragraph{HIL Capabilities and Insights}
The high-throughput, low-latency configuration demonstrated the real-world feasibility of executing a complete visual feedback control loop onboard. HIL testing allowed the team to observe discrepancies between simulated sensor characteristics and physical outputs, enabling fine-grained adjustments in the detection module. Importantly, this phase validated not only the algorithmic correctness but also the hardware-software compatibility of the end-to-end system, serving as a critical readiness gate for real-world deployment.

\paragraph{Key Challenges in HIL Testing}
While Hardware-in-the-Loop (HIL) testing offers critical advantages in evaluating the integration of perception, planning, and control modules on real hardware, several challenges remain that limit its fidelity and scalability. A major issue lies in maintaining the stringent real-time performance required for advanced drone maneuvering. In practice, HIL environments rely on frameworks such as ROS and Unreal Engine, which may not consistently support millisecond-level control loop execution \cite{TomLooman}. During our own HIL testing involving AirSim integrated with PX4 and ROS, noticeable latency was observed within the control module. This latency revealed synchronization and computational bottlenecks that could impact closed-loop stability and responsiveness in time-sensitive operations.

Another persistent challenge in HIL testing is the realistic modeling of complex environmental dynamics. Single-simulator environments often struggle to emulate fine-grained physical phenomena such as turbulent wind gusts, light reflections, or GPS signal degradation \cite{zhang2024dronewis}. These limitations can lead to mismatches between simulation and field results, particularly when testing robustness against external disturbances or failure recovery mechanisms.

Furthermore, current simulations typically employ generic or simplified sensor and actuator models. The simulated sensor noise profiles often do not precisely match those observed in actual drone hardware, potentially leading to discrepancies during sensor fusion and perception tasks. For example, minor timing drifts or sensor miscalibrations common in real sensors may not be captured in simulation, causing optimistic evaluations during testing. Similarly, actuator dynamics including latency, saturation effects, and mechanical wear are often idealized in simulation. This abstraction fails to represent the full complexity and unpredictability of physical drone responses under real-world operational conditions.

In summary, while HIL testing plays a vital role in bridging the gap between simulation and field deployment, its current limitations highlight the need for next-generation testing paradigms. Incorporating digital twins, co-simulation platforms, and more precise modeling of real-world noise and actuator behaviors can substantially increase the realism, reliability, and scalability of HIL-based validation pipelines.

\subsection{Controlled Real-World Testing}
Real-world testing constitutes the final and most critical stage in the autonomous drone testing pipeline. While earlier testing stages such as Software-in-the-Loop and Hardware-in-the-Loop rely primarily on simulated scenarios and controlled hardware setups, real-world testing uniquely exposes the drone system to the full complexity and unpredictability of physical environments. This stage allows developers to comprehensively validate the drone's end-to-end system performance under realistic operational conditions, capturing interactions with dynamic and unpredictable elements, such as moving obstacles, variable lighting conditions, and changing weather patterns. Furthermore, real-world testing assesses the long-term stability and robustness of the drone under extended operation scenarios, ensuring system resilience. Critically, this stage also serves to verify that the drone adheres strictly to applicable regulatory requirements and safety standards, which is essential prior to actual deployment.

While the preceding simulation-based and hardware-integrated testing phases provide valuable insights into system correctness, responsiveness, and reliability, certain integration issues and environment-induced challenges can only surface during real-world testing. Thus, despite its higher logistical complexity and potential safety risks, real-world testing is indispensable for ensuring that the drone operates safely and effectively under authentic operational conditions.

The initial phase of real-world testing is typically performed within highly controlled indoor environments, such as motion capture rooms, drone cages, or large, empty indoor arenas. In these controlled settings, environmental factors like wind, lighting, and temperature can be carefully regulated or eliminated altogether, providing a safe and predictable testing scenario. Controlled indoor testing is particularly suitable for validating essential drone functionalities, including basic maneuvers such as take-off, landing, and waypoint following, before transitioning to less predictable outdoor conditions.

The benefits of controlled indoor testing are substantial. These environments ensure initial validation can be conducted safely, significantly reducing the risk of hardware damage or safety incidents. Moreover, the highly controlled conditions enable precise measurement and rigorous evaluation of flight accuracy, repeatability, and system reliability. Nevertheless, controlled indoor testing does carry notable limitations: it does not replicate the diverse and often challenging environmental variables encountered outdoors. Therefore, this stage typically precedes comprehensive outdoor tests to fully validate drone system performance under authentic operational conditions.

\subsubsection{Real-World Testing Preparation and Execution}
Conducting controlled real-world testing requires careful planning and meticulous preparation to ensure both safety and meaningful evaluation outcomes. Prior to commencing any tests, the drone hardware and software must undergo rigorous pre-flight checks, including sensor calibration, battery status verification, and communication link testing. A clearly defined test plan detailing mission objectives, safety protocols, expected behaviors, and contingency measures should be established and communicated among all team members. Additionally, appropriate testing locations such as enclosed indoor environments with safety nets or controlled outdoor facilities must be selected to ensure compliance with regulatory guidelines and minimize risks. The execution phase involves progressively validating individual drone functionalities, systematically increasing complexity until complete end-to-end operations, including marker-based landing and dynamic obstacle avoidance, are thoroughly assessed under realistic conditions.

\paragraph{Risk Assessment and Safety Planning}
The first and most crucial step in controlled real-world testing is performing a thorough risk assessment and developing a comprehensive safety plan. Before any flight activity, it is essential to clearly define the testing objectives such as validating a landing algorithm or verifying waypoint tracking accuracy and establish exit criteria to determine when a test should be considered complete, successful, or prematurely terminated due to safety concerns.

A formal hazard analysis should be conducted to identify potential risks associated with drone malfunction, sensor failure, software bugs, or human error. For each identified risk, appropriate mitigation strategies must be developed. These may include physical safeguards such as tethering the drone, procedural controls like pre-flight checklists, and environmental controls such as netted enclosures or indoor geofencing boundaries.

In addition, emergency recovery procedures must be prepared in advance and rehearsed by all team members. This includes the deployment of manual override mechanisms such as a kill switch to immediately cut power in the event of unsafe behavior and the use of geofencing tools to prevent the drone from leaving the designated test area. These proactive safety measures are essential to minimize harm to people, property, and the vehicle itself, ensuring that controlled real-world testing proceeds in a secure and responsible manner.

\paragraph{Test Site Preparation}
Selecting and preparing an appropriate test site is essential for ensuring that controlled real-world testing is both effective and compliant with all relevant safety and legal requirements. The choice of test location should be informed by the complexity of the testing scenario and the associated risk level. For early-stage testing of basic maneuvers or system integration, a fully enclosed indoor facility such as a warehouse, motion capture lab, or drone cage offers a safe and controllable environment. In contrast, more complex or longer-range scenarios may require larger indoor arenas or partially controlled outdoor areas.

Regulatory compliance is a critical aspect of site preparation. While indoor tests generally fall outside civil aviation regulations, institutional protocols may still require completing formal documentation such as a Job Safety Assessment (JSA). For outdoor tests, it is important to comply with local aviation authority guidelines, such as those issued by the Civil Aviation Safety Authority (CASA) in Australia. This may involve obtaining necessary flight permissions, filing Notices to Airmen (NOTAMs), and ensuring that all pilots and observers are certified and briefed.

Additionally, the test site must be equipped with the necessary infrastructure to support testing operations. This can include indoor tracking systems (e.g., VICON or OptiTrack) for ground truth measurements, safety nets and barriers to prevent unintended flyaways, observation stations for team monitoring, and reliable communication links between the drone and the ground station. Preparing the environment in this way ensures a safe, measurable, and controlled setting in which meaningful real-world evaluations can take place.

\paragraph{Pre-Flight System Validation}
Before initiating any flight in a controlled real-world testing environment, it is essential to perform comprehensive pre-flight validation to confirm that all subsystems are functioning correctly. This begins with routine ground checks to ensure that hardware components are properly connected and securely mounted. Critical items such as battery levels, motor connections, propeller integrity, and payload attachments must be inspected. On the software side, it is important to verify that configuration parameters such as mission profiles, control gains, and flight modes are correctly loaded and consistent with the intended test plan.

A reliable GPS lock is essential for outdoor tests, and its status should be monitored to ensure stable satellite reception before arming the drone. Similarly, all onboard sensors (e.g., IMU, barometer, magnetometer, and cameras) must be correctly calibrated to prevent navigation errors during flight. Communication links whether over telemetry, Wi-Fi, or Ethernet between the drone, onboard computing platform, and ground control station should also be tested for stability and low latency.

Where possible, performing a dry-run simulation of the planned flight can help identify logic errors or unexpected behaviors in mission scripts or control algorithms. This step, often performed with propellers removed or in simulation replay mode, provides an additional layer of assurance that the drone system is ready for safe and effective operation during the physical test. Collectively, these pre-flight validations are crucial for minimizing operational risk and ensuring mission reliability.

\paragraph{Test Execution and Monitoring}
Once pre-flight validations are complete, the real-world testing process proceeds with the execution of planned test scenarios under close observation. To ensure safety and reduce the risk of unexpected failures, tests should begin with low-risk scenarios such as basic take-off and hover maneuvers before gradually progressing to more complex operations like trajectory following, obstacle avoidance, or autonomous landing. This incremental approach allows the development team to evaluate subsystem performance and integration fidelity in a controlled manner.

During execution, it is critical to continuously monitor the drone’s behavior through a ground control station (GCS) interface, which provides real-time telemetry, sensor feedback, flight mode status, and visual cues. Any deviations from expected behavior such as instability, incorrect position tracking, or delayed responses should trigger immediate pause or termination procedures as defined in the safety plan.

Equally important is the comprehensive recording of all relevant data streams throughout the test. This includes onboard sensor readings (e.g., IMU, GPS, barometer, cameras), control inputs and setpoints, actuator outputs, and full telemetry logs. Tools such as \texttt{rosbag}, PX4’s \texttt{ulog}, and real-time dashboards can be used to capture and archive this data. These records are essential for post-test analysis, helping developers identify root causes of anomalies, verify performance metrics, and refine system parameters for future iterations.

\paragraph{Post-Test Analysis}
After each controlled real-world test, a thorough post-test analysis is essential to evaluate system performance, identify anomalies, and guide further development. The first step involves reviewing the recorded flight data such as sensor logs, control commands, flight trajectories, and telemetry streams to detect any irregularities or unexpected behaviors during the test. Common issues might include sensor noise spikes, unstable control responses, communication dropouts, or deviations from intended paths.

Once the data is parsed, outcomes should be systematically compared against predefined success criteria and safety thresholds. For instance, metrics such as landing precision, waypoint accuracy, response latency, or energy consumption can be assessed relative to baseline expectations. Any violations of safety limits (e.g., altitude overshoots, excessive drift, delayed emergency stops) should be flagged for immediate correction.

Based on the findings, necessary iterations can be planned. This may involve software-level bug fixes, refining algorithm parameters (e.g., control gains or filtering thresholds), updating mission logic, or recalibrating hardware components. Post-test analysis not only provides actionable insights for resolving current issues but also enhances system robustness for future, more complex deployments. Ultimately, this feedback loop transforms raw test outcomes into targeted engineering improvements, accelerating the development of a safe and reliable autonomous drone system.

\subsubsection{Case Study: Controlled Real-World Testing for Marker-Based Autonomous Landing System}
Following comprehensive validation through Software-in-the-Loop and Hardware-in-the-Loop stages, the marker-based autonomous landing system advanced to controlled real-world testing to assess its capabilities in realistic yet risk-mitigated environments.

\begin{figure}[!ht]
	\centering
	\includegraphics[width=0.4\linewidth]{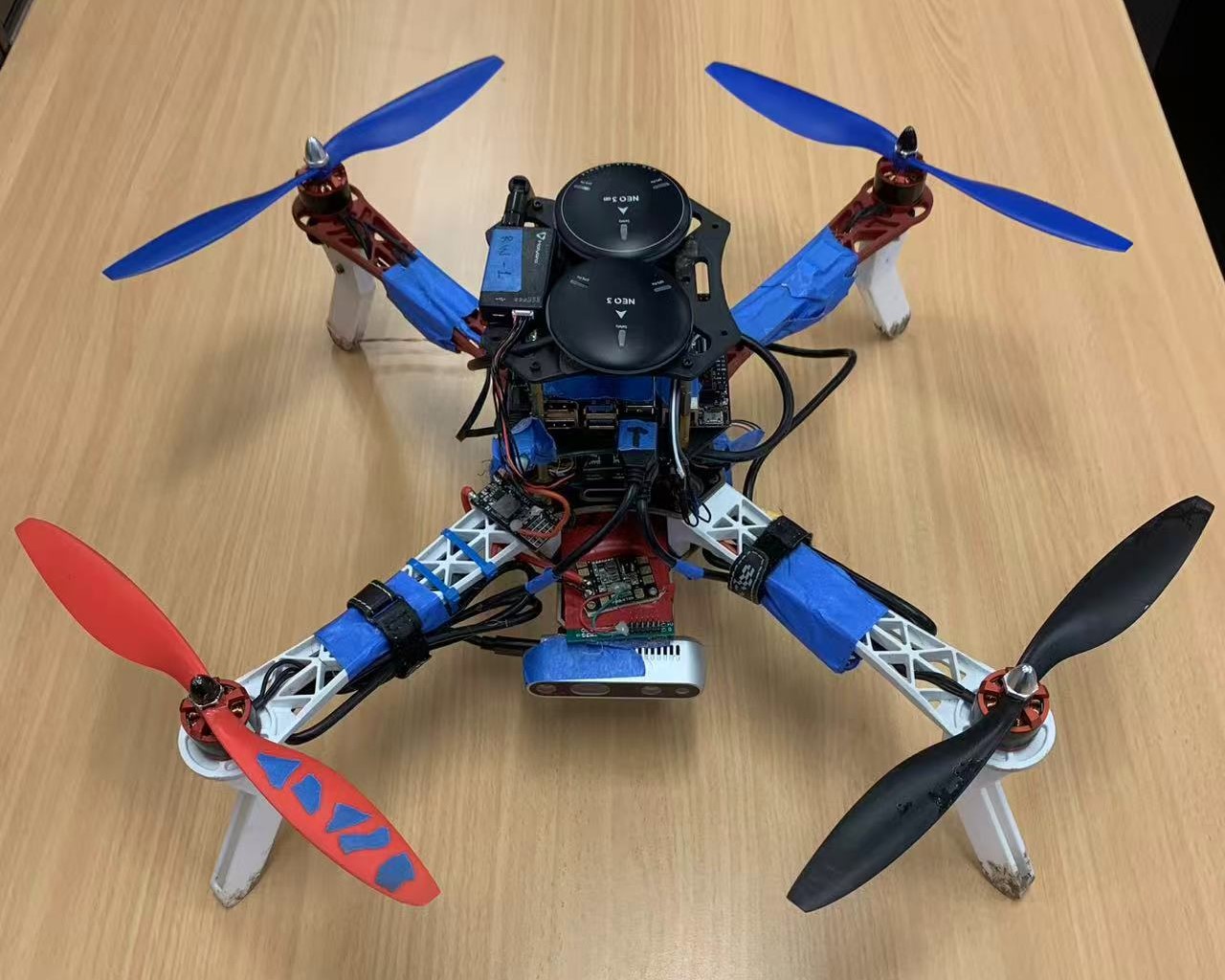}
        \caption{Fully assembled autonomous drone platform.}
	\label{cots}
\end{figure}

The tested system comprised an integrated pipeline consisting of a visual marker detection module, a trajectory planner, and a precision descent controller. The primary objective of the controlled real-world testing phase was to validate this end-to-end autonomous landing system under realistic physical dynamics, while maintaining a high degree of environmental control. To enable rapid prototyping and iterative testing, the hardware platform was built using standard commercial off-the-shelf (COTS) components as shown in Figure~\ref{cots}. The aerial platform was based on an F450 quadcopter frame \cite{F450}, equipped with a Jetson Nano \cite{JetsonNano} (4 GB) as the onboard companion computer. A CUAV X7+ Pro Pixhawk flight controller \cite{CUAV} was used to ensure robust flight performance and enhanced sensor interfacing. The perception system was composed of a forward-facing Intel RealSense D435 \cite{D435} for obstacle detection and a downward-facing RealSense D435i for visual marker tracking. For positioning and altitude estimation, the system employed a CUAV NEO 3 GPS module \cite{NEO3} and a TFmini Plus LiDAR ranging sensor \cite{TFmini}, respectively. This modular setup allowed reliable evaluation of the perception and control stack in a controlled indoor setting while enabling rapid hardware modifications throughout the test campaign.

\begin{figure}[!ht]
    \centering
    \subfloat[Autonomous drone tethered with a safety line.]{\includegraphics[width=5cm]{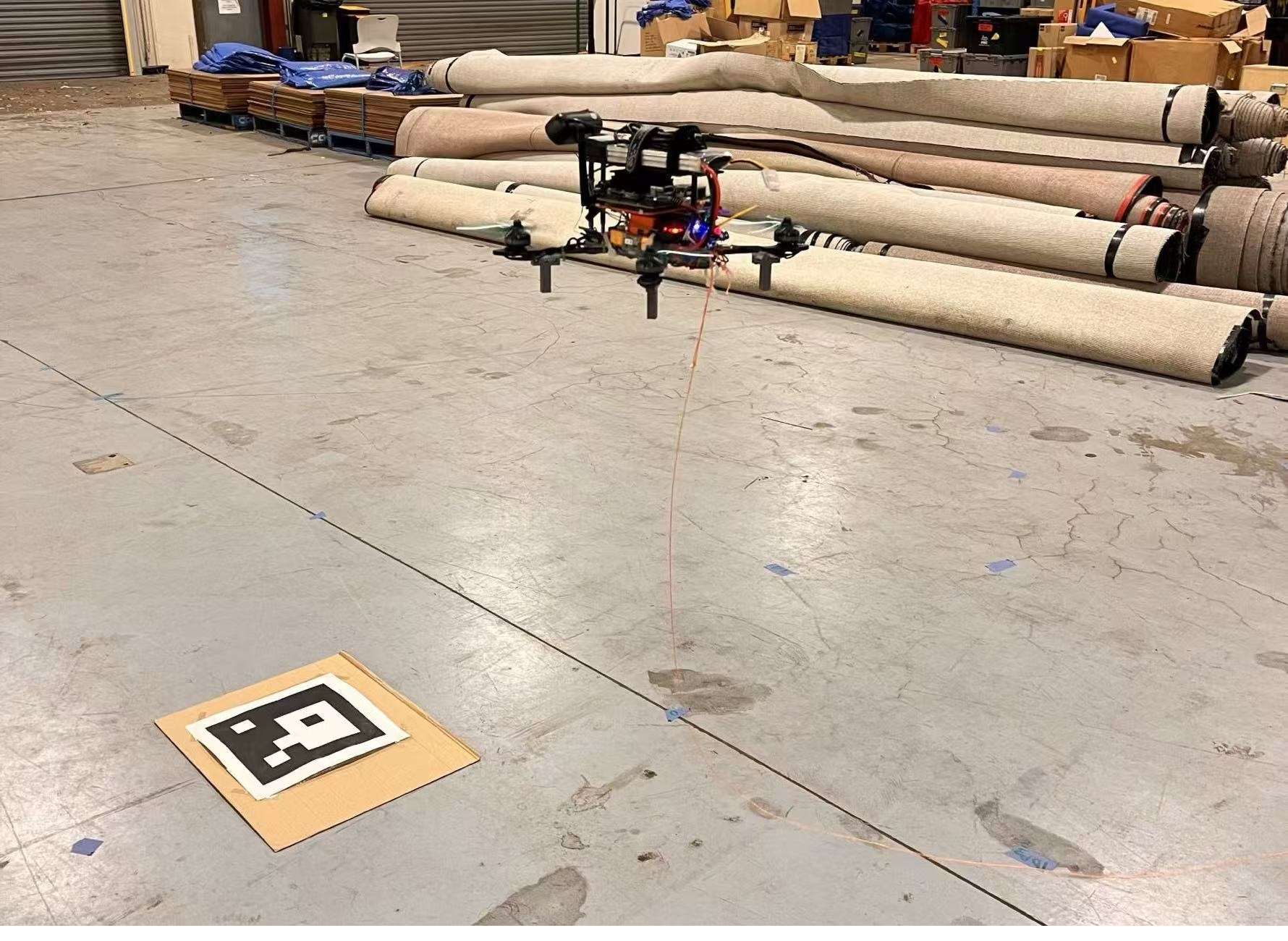}
    \label{fig:tether}}
    \hspace{1cm}
    \subfloat[ArUco marker and lightweight physical obstacle.]{\includegraphics[width=5cm]{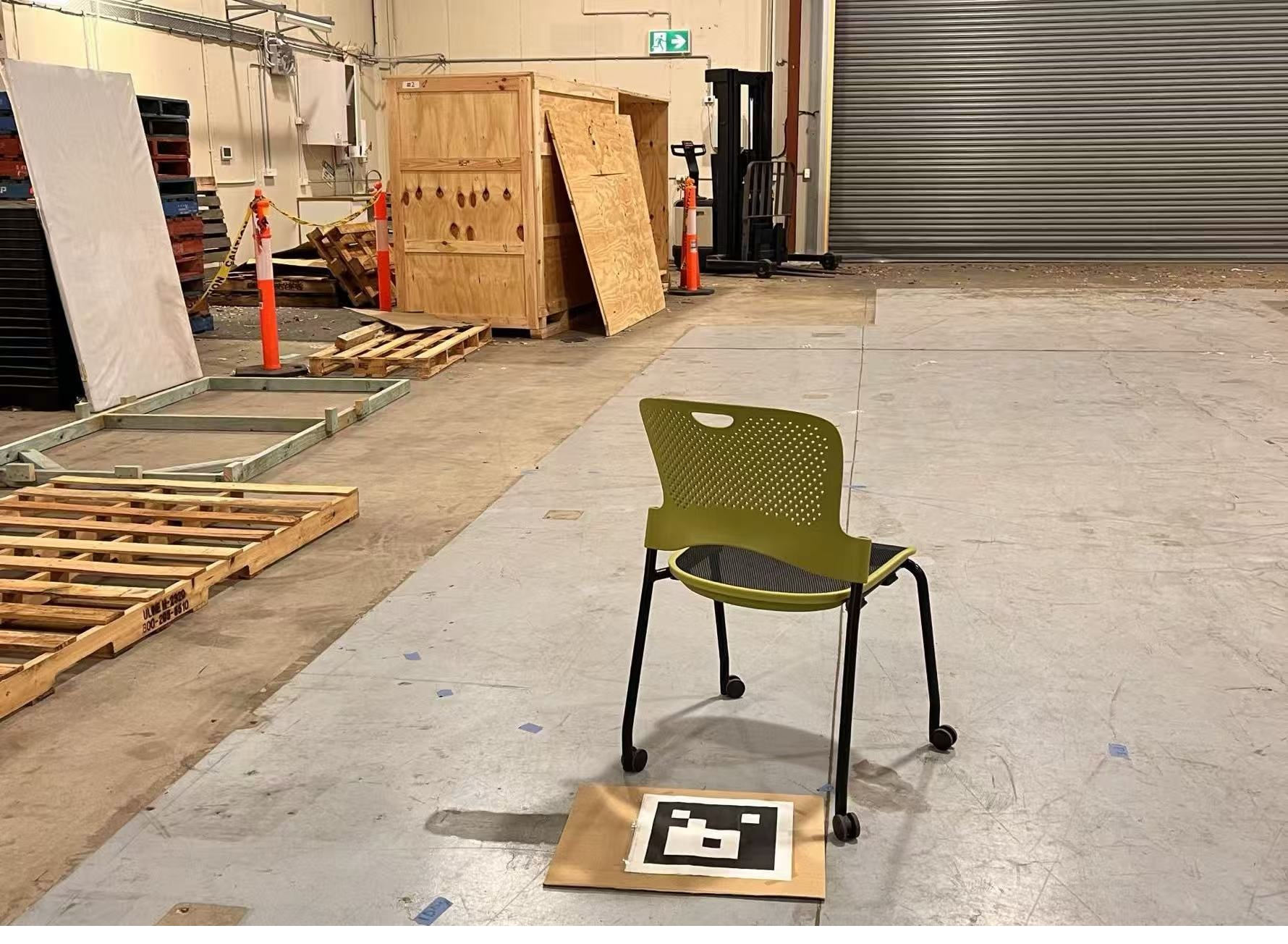}
    \label{fig:aruco}}
    \captionsetup{font=small}
    \caption{Controlled indoor testing setup for validating the marker-based landing system.}
    \label{fig:warehouse}
\end{figure}

Testing was conducted in a large indoor warehouse equipped with safety infrastructure including physical barriers and clearly marked observation zones. To minimize risk during early-stage flights, the drone was tethered with a safety line, allowing emergency retrieval while maintaining sufficient freedom of motion for realistic testing. A series of ArUco markers were placed at known ground locations, and lightweight physical obstacles were introduced to simulate occlusions and test robustness of the perception and recovery logic. Figure~\ref{fig:tether} shows the drone operating in the warehouse environment, with tether lines visibly attached for safety and control. Figure~\ref{fig:aruco} illustrates an ArUco marker on the landing pad and nearby lightweight obstacles used to test occlusion tolerance. These images collectively document the realism, control, and safety-conscious design of the test setup.

The mission plan involved autonomous take-off, marker detection during hovering, trajectory alignment, and smooth vertical descent. The ground station was used to monitor telemetry and live video feeds, and the entire test environment was recorded for post-flight analysis.

The system successfully achieved precision landings, consistently reaching within ±5 cm of the marker center across multiple trials. The landing pipeline demonstrated reliable descent behavior even under moderate visual occlusion and dynamic lighting conditions. Additionally, indoor flights allowed controlled testing of edge-case scenarios such as partial marker loss and abrupt reappearance, which revealed limitations in the initial recovery heuristics.

These real-world observations led to several key refinements in the landing system. First, the perception module was enhanced through improved temporal filtering, which helped reduce false negatives in marker detection caused by brief occlusions or motion blur. Second, the descent controller was adjusted to apply trajectory smoothing techniques, which significantly improved landing stability, particularly when operating near obstacles or under turbulent conditions. Lastly, an emergency handling mechanism was introduced, enabling the drone to hover and initiate a search routine if the landing marker became temporarily invisible during descent. These iterative improvements collectively increased the robustness and reliability of the overall system in complex environments.

This phase of controlled real-world testing significantly enhanced the system’s robustness. It bridged the gap between simulation accuracy and real-world variability, providing crucial insights that were not fully uncovered during SIL and HIL stages. Furthermore, the recorded flight logs, images, and video datasets served as high-quality material for downstream regression testing and retraining of AI-based perception modules.

\subsection{In-Field Testing}
After simulation-based evaluation, software-in-the-loop refinement, hardware-in-the-loop integration, and controlled real-world testing, the final phase in the drone testing pipeline is in-field testing. This stage subjects the autonomous drone to unconstrained, natural environments where real-world variables such as wind, lighting, terrain diversity, and human presence are no longer abstracted or simulated but encountered in full complexity.

Unlike indoor or laboratory-controlled tests, in-field testing is essential for validating system behavior in realistic deployment scenarios, such as autonomous package delivery, search and rescue missions, or environmental monitoring. It also plays a critical role in verifying long-term system robustness, safety protocol effectiveness, and regulatory compliance.

To balance operational realism with safety and compliance, in-field testing is typically conducted in progressive stages, beginning with controlled outdoor evaluations before advancing to fully public environments. Each stage introduces increasing levels of environmental uncertainty and legal responsibility, necessitating careful planning, monitoring, and contingency design.

\subsubsection{Controlled Outdoor Testing}
The first stage of in-field testing involves deployment in private or semi-private outdoor environments, such as university campuses, agricultural test fields, or fenced-in drone research parks. These areas offer natural environmental exposure such as wind, variable GPS signal quality, and changing ambient light while maintaining a degree of spatial control and regulatory insulation.

Prior to testing, teams conduct site inspections, file safety protocols, and obtain required local or university-level permissions. Tethering systems, visual tracking tools, and automated geofencing are commonly employed as safety mechanisms to prevent flyaways or unintended intrusion into restricted airspace. Safety pilots or remote observers may also be stationed to manually intervene if anomalies occur.

This testing phase is particularly suitable for evaluating mission autonomy, perception under dynamic lighting, outdoor marker tracking, or adaptive control algorithms. It also allows iterative tuning of system parameters with relatively low risk.

Benefits of controlled outdoor testing include a good balance between environmental realism and experimental control, and a safe platform for validating advanced autonomy logic before facing public exposure. However, limitations remain: the weather dependency, lack of bystander variability, and limited environmental unpredictability constrain the completeness of test coverage.

\subsubsection{Public Environment Testing}
The final and most demanding stage of in-field testing takes place in open public or semi-public environments, such as urban parks, lakesides, or suburban streets, typically under strict legal and regulatory compliance. This includes securing flight permissions from aviation authorities, notices to airmen (NOTAMs), and adherence to airspace classification rules (e.g., CASA in Australia or FAA in the U.S.).

In this setting, drones are exposed to real-world conditions, including the presence of people, vehicles, dynamic obstacles, unpredictable weather, and interference sources. For instance, autonomous delivery drones may be tested in suburban neighborhoods to validate navigation, obstacle avoidance, human interaction protocols, and system redundancy.

The benefits of this stage are substantial: it provides full-scale validation in the intended deployment environment, enabling comprehensive characterization of system performance, reliability, and user safety. It is often a prerequisite for regulatory certification and commercial deployment.

However, public environment testing is high-risk, requiring rigorous planning, extensive fallback procedures, redundant communication links, and automated emergency recovery protocols such as return-to-home or safe-landing behaviors. Tests are usually supported by real-time ground control stations and supervised by cross-functional safety teams.

In summary, in-field testing completes the drone testing lifecycle by verifying that the system can perform reliably and safely in its intended operational context, thus bridging the final gap between lab readiness and field deployment.

\subsubsection{Case Study: In-Field Testing for Marker-Based Autonomous Landing System}
Following successful controlled indoor validations, the marker-based autonomous landing system advanced to controlled outdoor in-field testing conducted at a designated university test field. This stage focused on evaluating the drone's marker-based landing capabilities under realistic environmental conditions, including variability in GPS signals, wind disturbances, and natural lighting fluctuations. Such outdoor tests were essential to assess system robustness and readiness for potential real-world deployment scenarios.

\paragraph{Experimental Setup and Procedures}
Testing at this stage required comprehensive planning and adherence to strict regulatory compliance procedures. All team members involved in flight operations held valid flight registration numbers issued by Australia's Civil Aviation Safety Authority (CASA). For larger flight activities, the unmanned aerial vehicles (UAVs) were formally registered under the appropriate individual aviation reference numbers, in accordance with CASA guidelines. Prior to each flight, detailed information such as drone specifications, downward-facing camera details, flight date, planned locations, maximum flight altitude (limited to 10–12 meters due to proximity to helicopter traffic zones), and a thorough flight plan was provided to the university's flight safety officer for approval. Additionally, a comprehensive Job Safety Assessment (JSA) was completed, documenting potential hazards and corresponding mitigation strategies in alignment with CASA regulations. Each flight session was staffed by a qualified pilot and at least one dedicated observer to ensure safety and rapid emergency response capability.

\paragraph{Test Execution and Outcomes}
Outdoor tests primarily aimed to evaluate landing precision, stability, and robustness against external disturbances such as wind gusts and intermittent GPS quality. The drone performed autonomous missions involving visual marker detection, trajectory planning, and controlled descents onto ground-placed ArUco markers. Tests revealed minor deviations in landing accuracy, with observed errors averaging approximately ±15 centimeters from the marker center. Detailed analysis indicated these deviations primarily resulted from wind-induced disturbances during the final descent stage.

\paragraph{Challenges and Solutions}
One notable challenge encountered was the limited robustness of the original landing control algorithm under unpredictable wind gust conditions, leading to slight positional drift. To address this, the control algorithm was retuned to adopt more aggressive stabilization parameters, significantly improving the drone's resistance to environmental perturbations. Additionally, outdoor scenarios highlighted scenarios where intermittent GPS signals and variable lighting conditions affected marker detection reliability. In response, the perception algorithm parameters were refined to handle rapid changes in lighting, and fallback logic was enhanced for maintaining stability and positioning accuracy even during brief marker losses.

\paragraph{Lessons Learned and Benefits}
These outdoor in-field tests provided invaluable insights that were not fully observable during earlier indoor phases. By exposing the system to natural environmental elements, the team successfully identified and addressed subtle integration issues and control inadequacies. These iterative refinements greatly improved overall system resilience, significantly enhancing landing accuracy, stability, and safety confidence. The extensive datasets collected during these trials also serve as valuable resources for future regression testing, scenario simulations, and training of advanced perception algorithms, solidifying the foundation for subsequent full-scale public environment testing and eventual real-world operational deployment.

\section{Future Trends in Drone Testing}
As autonomous drone systems grow in complexity and are deployed in increasingly dynamic and safety-critical environments, traditional testing pipelines are facing new limitations. While SIL, HIL, and real-world testing collectively offer a comprehensive validation framework across different abstraction layers, our experience in practical deployment reveals several persistent limitations. These traditional methods, though indispensable, are often insufficient in capturing rare edge cases, modeling complex environments with high fidelity, or supporting continuous learning and adaptation. Through our implementation and iterative testing of autonomous drone systems, we have identified key challenges that point to the need for more intelligent, adaptive, and scalable testing paradigms.

To address these emerging challenges, the drone testing community is shifting toward more intelligent and system-aware testing frameworks that go beyond traditional staged pipelines. In particular, there is growing interest in integrating AI-driven methods for test generation and scenario prioritization, leveraging co-simulation environments that bridge multiple physical and cyber domains, and constructing digital twin systems that closely mirror the physical characteristics and runtime behaviors of deployed drones. These trends represent a departure from isolated testing stages toward a more continuous, adaptive, and high-fidelity validation ecosystem, better aligned with the operational complexity of next-generation autonomous systems.

To better understand how emerging techniques are addressing the limitations of traditional testing approaches, we summarize the key motivations, core capabilities, and typical use cases of three major trends that will shape the evolution of autonomous drone testing. Table~\ref{Emerging_Trends} provides a comparative overview of these trends, setting the stage for a more detailed discussion in the subsequent sections. As shown in the table, one of the most promising developments lies in the integration of neurosymbolic reasoning and large language models. These techniques aim to automate and enrich the testing process by combining structured logic with data-driven generalization, which we explore in detail in the following section.

\begin{table*}
	\centering
	\caption{Emerging Trends in Autonomous Drone Testing}
	\label{Emerging_Trends}
	\begin{threeparttable}
		\renewcommand\arraystretch{1.5}
		\newcolumntype{A}{>{\raggedright}p{0.25\textwidth}}
		\newcolumntype{B}{>{\raggedright\arraybackslash}p{0.45\textwidth}}
		\newcolumntype{C}{>{\raggedright\arraybackslash}p{0.51\textwidth}}
		\newcolumntype{D}{>{\raggedright\arraybackslash}p{0.35\textwidth}}
		\scalebox{0.6}{
			\begin{tabular}{A B C D}
				\toprule
				\textbf{Trend} & \textbf{Motivation} & \textbf{Key Capabilities} & \textbf{Representative Use Cases} \\ 
				\midrule 
				\makecell[l]{Neurosymbolic \& \\ LLM-based Testing} & \makecell[l]{Manual test creation lacks \\ semantic coverage and adaptivity} & \makecell[l]{- Intelligent test generation \\ - Reasoning over mission constraints \\ - Scenario prioritization using learned models} & \makecell[l]{- Corner-case detection \\ - Task-specific robustness testing \\ - Automated scenario design} \\
				\midrule
				\makecell[l]{Co-simulation \\ Environments} & \makecell[l]{Single simulators (e.g. AirSim, \\ Gazebo) cannot capture \\ cross-domain dynamics} & \makecell[l]{- Multi-agent \& multi-domain interaction \\ - Modular simulator integration \\ ~~(e.g. physics + network) \\ - Synchronization across subsystems} & \makecell[l]{- Urban air mobility testing \\ - Vehicle-drone interaction \\ - Network-aware UAV operations} \\
				\midrule
				\makecell[l]{Digital Twin \\ Simulation} & \makecell[l]{Traditional simulation lacks alignment \\ with real hardware dynamics} & \makecell[l]{- High-fidelity real-world mirroring \\ - Runtime synchronization \\ - Predictive diagnostics \& update testing} & \makecell[l]{- Mission rehearsal \\ - Runtime anomaly detection \\ - Safe firmware validation} \\
				\bottomrule
		\end{tabular}}
	\end{threeparttable}
\end{table*}

\subsection{Integration of Neurosymbolic and Large Language Models}
As autonomous drone systems grow more sophisticated and increasingly operate in complex, unstructured environments, there is a pressing need to improve the intelligence, precision, and reliability of testing frameworks. Traditional approaches whether blackbox testing for machine learning models or logic-based testing for rule-driven systems often fall short in capturing semantically nuanced behaviors or edge-case interactions. This has led to growing interest in neurosymbolic and LLM-based testing paradigms, which offer a promising pathway to unify data-driven perception with symbolic reasoning under a common framework.

To bridge the longstanding divide between neural and symbolic methods, the emerging neurosymbolic paradigm integrates the perception and pattern recognition capabilities of deep learning with the logical structure, rule-based reasoning, and formal constraints of symbolic systems. A representative example is the NeuroStrata \cite{Zheng2025NeuroStrataHN} framework , which proposes a layered architecture for integrating large language models (LLMs), symbolic constraints, and differentiable reasoning modules throughout the cyber-physical system software stack. In this framework, LLMs are employed to generate semantically rich test cases, conditioned on mission specifications, environmental context, and prior incidents. These generated tests may incorporate high-level goals such as “landing in dense urban zones,” combined with dynamic environmental cues like GPS anomalies, intermittent signal loss, or nearby moving objects. This is achieved by combining LLMs with logic rules to generate complex, semantically rich scenarios (e.g., GPS degradation with moving obstacles), ensuring that the test inputs reflect realistic, mission-relevant challenges. To ensure operational feasibility and safety, these LLM-generated test scenarios are then constrained using symbolic logic. Constraints can include airspace rules, power consumption limitations, mission-critical priorities, or geofencing boundaries. This dual-process ensures that test cases are not only creative and comprehensive but also realistic and executable within system capabilities.

NeuroStrata also introduces mechanisms for differentiable program induction and symbolic execution over these LLM-generated programs. This enables test agents to explore both syntactic variability and deeper semantic behaviors that align with mission requirements and safety properties. For instance, a neurosymbolic test case could simulate a situation like “an emergency landing triggered by visual marker loss and GPS spoofing during final descent,” allowing engineers to probe the system’s resilience in edge-case operational scenarios. Such detailed testing goes beyond traditional fuzzing or manually scripted validation. It enhances the coverage and reasoning ability of the testing system and enables the discovery of rare but critical failure modes. Importantly, the integration of neurosymbolic paradigms brings a degree of semantic understanding, test automation, and formal reasoning that is essential for advancing toward certification-grade validation of autonomous drone systems.

In conclusion, neurosymbolic and LLM-driven testing strategies have the potential to transform the validation pipeline for autonomous drone testing. By capturing both abstract mission logic and real-world environmental interactions, these approaches can systematically validate behaviors that would otherwise go untested, especially in safety-critical applications such as autonomous landing, obstacle avoidance, and emergency rerouting.

\subsection{Co-simulation Environments}
As autonomous drone systems grow in complexity, conventional single-simulator setups are often insufficient to capture the nuanced behaviors and interactions encountered in real-world missions. Limitations in standalone simulators such as simplified agent behaviors, limited environmental physics modeling, and poor scalability to multi-agent scenarios have prompted a shift in testing methodologies. To address these gaps, emerging trends suggest co-simulation environments that link real hardware with software models of weather, RF propagation, or swarm coordination, creating comprehensive, modular platforms for high-fidelity experimentation.

Co-simulation environments are designed to integrate multiple specialized simulation tools, each optimized for a specific domain such as aerodynamics, network communications, sensor modeling, or real-time physics, into a cohesive and synchronized testbed \cite{gomes2018co}. Typical implementations may combine Unreal Engine \cite{UnrealEngine} (for photorealistic rendering), OMNeT++ \cite{varga2008overview} (for communication network simulation), and MATLAB/Simulink \cite{Simulink} (for advanced control logic and physical modeling). This enables highly precise, multi-domain modeling of operational scenarios that mirror real-world complexities such as turbulent airflow, GPS spoofing, RF interference, or multi-agent interactions in dense urban airspaces.

Furthermore, co-simulation frameworks now leverage AI-driven tools for scenario generation, test orchestration, and environment adaptation, extending their capabilities beyond fixed scripted tests. This dynamic environment control allows developers to explore edge-case behaviors, automate the discovery of rare failure modes, and generate realistic synthetic datasets to augment AI training pipelines. For example, one representative application involves the integration of Simulink-designed flight controllers with AirSim. Here, control algorithms are prototyped in Simulink, automatically compiled into C code for PX4 flight firmware, and deployed within the AirSim simulation environment to test performance under extreme weather or terrain conditions. In another workflow, AirSim-generated visual data, simulating diverse environmental lighting and weather patterns, can be used directly with MATLAB’s Deep Learning Toolbox to train and validate object detection models like TPH-YOLOv5 \cite{zhu2021tph}, enabling efficient AI pipeline development.

Co-simulation environments also enable scalable validation for large-scale mission types, including urban air mobility, autonomous delivery networks, and disaster response operations. These scenarios demand synchronized modeling of drones, ground agents, infrastructure constraints, and environmental effects, which co-simulation platforms are well-positioned to deliver. Through their modular, extensible architecture, such environments allow for rapid testing iterations, more accurate verification of control and perception algorithms, and comprehensive safety assessments under diverse real-world analogues.

In summary, co-simulation frameworks represent a significant evolution in drone testing infrastructure. By combining the flexibility of modular simulation platforms with the depth of domain-specific modeling, they provide the fidelity, realism, and scalability essential for validating next-generation autonomous systems, bridging the gap between simulation and real-world deployment with unprecedented accuracy.

\subsection{Digital Twin Enabled High-Fidelity Simulation}
As drone autonomy continues to advance, there is a growing demand for simulation frameworks that can not only replicate environmental interactions but also mirror the specific hardware and software characteristics of individual drone platforms. Traditional simulation environments like AirSim and Gazebo have proven invaluable for early-stage development and algorithm prototyping. However, as highlighted in previous sections, they often rely on generalized models and pre-programmed abstractions that fall short when evaluating nuanced behaviors, system degradations, or hardware-specific dynamics.

To address this gap, the concept of Digital Twins \cite{rasheed2020digital} is gaining prominence within drone testing communities. A digital twin is defined as a highly accurate virtual replica of a specific physical drone system, continuously synchronized and calibrated against real-time data collected from its physical counterpart. Unlike traditional simulators, digital twins incorporate detailed, system-specific information including real sensor calibration data, precise actuator response characteristics, and comprehensive maintenance and mission logs into their virtual representations.

Simulation platforms such as AirSim play a foundational role in digital twin implementation, providing realistic visualization environments built upon engines like Unreal, thus forming the visual and interactive interface for the digital twin. Furthermore, advanced data acquisition technologies including IoT sensors and high-speed communication networks (such as 5G) enable continuous streaming of real-world sensor measurements directly into the simulation model \cite{NVIDIADigitalTwin}. This integration allows the digital twin to maintain accurate synchronization with the physical drone, significantly enhancing predictive capabilities and operational insights.

Digital twins leverage simulation tools not only as visualization interfaces but also as comprehensive test and validation environments. For instance, while historical operational data is critical for training digital twin models, it is often limited or lacks coverage of rare edge cases. Simulation tools can generate synthetic data scenarios such as extreme weather events or component failures to bridge these gaps \cite{tao2019digital}. By providing diverse, high-quality synthetic datasets, AirSim enables training and validation of sophisticated AI-driven decision-making modules within the digital twin. Moreover, integrating AirSim with flight control software like PX4 facilitates seamless Hardware-in-the-Loop interactions, enabling robust virtual-to-physical testing and validation.

An illustrative example of the digital twin approach in drone testing is its application to autonomous landing algorithms. Future development roadmaps involve constructing detailed digital twins of drone platforms, allowing pre-mission validations under precisely replicated hardware characteristics and environmental parameters. In drone logistics applications, for instance, AirSim can virtually replicate drone flight trajectories and obstacle avoidance behaviors. During actual flights, real-time sensor data is continuously synchronized with the AirSim-based digital twin model, enabling predictive analysis such as estimating battery life \cite{TWAICE} or evaluating weather impacts \cite{HIGHTOPO} to optimize flight paths dynamically and enhance overall mission reliability.

As simulation tools evolve, their roles in drone testing are shifting from traditional offline validation toward integrated real-time components of digital twin frameworks. Platforms like AirSim already support real-time ROS data integration, enabling them to serve as interactive, synchronized components of live drone operations. Consequently, the clear boundary between offline simulation and real-time operational monitoring is gradually dissolving, giving rise to more holistic lifecycle management strategies for autonomous drone systems.

In summary, traditional simulation environments constitute the foundational building blocks, providing virtual modeling capabilities, algorithmic testing, and data generation infrastructures. The digital twin approach, however, represents a significant advancement, effectively elevating simulations by incorporating real-time data streams and closing the feedback loop between virtual models and physical deployments. This tight integration not only bridges the gap between theoretical design and practical operations but also provides powerful predictive and optimization capabilities across the entire operational lifecycle of autonomous drone systems.

\section{Conclusion}
Autonomous drones are rapidly transforming industries ranging from logistics and infrastructure inspection to environmental monitoring and emergency response. However, as these systems become more autonomous and are deployed in increasingly complex environments, the demand for robust, systematic, and safety-oriented testing pipelines becomes paramount. This paper presented a step-by-step guide to creating and implementing a comprehensive autonomous drone testing pipeline, progressing through the core testing stages: Software-in-the-Loop (SIL) simulation testing, Hardware-in-the-Loop (HIL) testing, controlled real-world testing, and in-field testing. Throughout each stage, we demonstrated how this pipeline was practically applied to the marker-based autonomous landing system, highlighting typical challenges and solutions encountered when developing real-world autonomous drone applications.

Beyond current best practices, this work also discussed future trends in drone testing, including: the integration of Neurosymbolic and Large Language Models to enable intelligent and automated test scenario generation; the emergence of co-simulation environments that unify heterogeneous simulators for more realistic and mission-relevant testing; and the advancement towards Digital Twin enabled simulation, offering ultra-high-fidelity, system-specific validation capabilities. Together, these innovations promise to make future drone testing pipelines more intelligent, automated, and aligned with the complexities of real-world deployment.

In conclusion, robust testing pipelines serve as critical infrastructure in autonomous drone development, underpinning the system’s ability to operate safely, reliably, and with public and regulatory confidence. By combining methodical multi-stage testing with forward-looking technologies, developers can ensure that autonomous drones meet the highest standards before entering the skies. The rapid evolution of autonomous drone capabilities necessitates a corresponding advancement in testing and validation methodologies to match their expanding operational scope and safety requirements. This work serves as both a practical guide and a forward-looking framework to support researchers, engineers, and practitioners in building the next generation of autonomous aerial systems with confidence.


\bibliographystyle{ACM-Reference-Format}
\bibliography{manuscript}

\newpage
\appendix

\section{Package Layout and Directory Overview}
\label{appendix:pkg-layout}
\begin{lstlisting}[language=bash]
 drone_simulation/
 |-- CMakeLists.txt
 |-- package.xml
 |-- launch/
 |   |-- drone_simulation.launch
 |   |-- drone_simulation_with_args.launch
 |   |-- marker_detector.launch
 |-- config/
 |   |-- params.yaml
 |   |-- marker_config.yaml
 |-- scripts/
 |   |-- marker_detector.py
 |   |-- auto_landing_controller.py
 |   |-- path_planner.py
 |-- msg/
 |   |-- LandingStatus.msg
 |-- srv/
 |   |-- TriggerLanding.srv
 |-- urdf/
 |   |-- drone_model.urdf.xacro
 |-- rviz/
 |   |-- drone_visualization.rviz
 |-- worlds/
 |   |-- indoor_warehouse.world
 |-- include/
 |   |-- drone_simulation/
 |       |-- some_header.hpp  # For C++ nodes
 |-- src/
     |-- some_cpp_node.cpp    # For C++ nodes
\end{lstlisting}

\begin{table}[!ht]
    \centering
    \caption{Explanation of Each Directory}
    \footnotesize
    \begin{tabular}{p{0.12\textwidth} p{0.6\textwidth}}
    \toprule
    \textbf{Directory} & \textbf{Purpose} \\
    \midrule
    \texttt{launch/} & Launch files for simulation startup, argument parsing, RViz, etc. \\
    \midrule
    \texttt{config/} & YAML files for loading ROS parameters (e.g., marker sizes, PID gains). \\
    \midrule
    \texttt{scripts/} & Python-based ROS nodes (e.g., perception, control, planning logic). \\
    \midrule
    \texttt{msg/} & Custom messages (e.g., \texttt{LandingStatus}) for inter-node communication. \\
    \midrule
    \texttt{srv/} & Custom services (e.g., \texttt{TriggerLanding}) to start behaviors on command. \\
    \midrule
    \texttt{urdf/} & Drone model (used for RViz or Gazebo visualization). \\
    \midrule
    \texttt{rviz/} & RViz configuration files for visualization setup. \\
    \midrule
    \texttt{worlds/} & Custom indoor or outdoor simulation worlds (if using Gazebo). \\
    \midrule
    \texttt{include/ src/} & Optional directories for C++ implementations. \\
    \bottomrule
    \end{tabular}
\end{table}

\newpage
\section{Launch File: \texttt{marker\_detector.launch}}
\label{appendix:marker-launch}
\begin{lstlisting}[language=xml]
<launch>
  <!-- Marker Detector Launch File -->
  <arg name="drone_name" default="drone_1"/>
  <arg name="camera_topic" default="/airsim_node/$(arg drone_name)/front_center_custom/Scene"/>
  <arg name="marker_pose_topic" default="/$(arg drone_name)/marker_pose"/>
  <arg name="marker_length" default="0.15"/>

  <!-- Example camera intrinsics (should be calibrated) -->
  <param name="camera_matrix" value="[[640, 0, 320], [0, 480, 240], [0, 0, 1]]" />
  <param name="dist_coeffs" value="[0, 0, 0, 0, 0]" />

  <!-- Launch marker detector node -->
  <node pkg="drone_simulation"
        type="marker_detector.py"
        name="marker_detector"
        output="screen"
        required="true">
    <param name="camera_topic" value="$(arg camera_topic)" />
    <param name="marker_pose_topic" value="$(arg marker_pose_topic)" />
    <param name="marker_length" value="$(arg marker_length)" />
  </node>
</launch>
\end{lstlisting}

\newpage
\section{Launch File: \texttt{drone\_simulation.launch}}
\label{appendix:simulation-launch}
\begin{lstlisting}[language=xml]
<launch>
  <!-- Arguments to configure drone behavior -->
  <arg name="drone_name" default="drone_1"/>
  <arg name="marker_length" default="0.15"/>
  <arg name="enable_visualization" default="true"/>

  <!-- Parameter values -->
  <param name="camera_matrix" value="[[640, 0, 320], [0, 480, 240], [0, 0, 1]]"/>
  <param name="dist_coeffs" value="[0, 0, 0, 0, 0]"/>

  <!-- Marker Detector Node -->
  <node pkg="drone_simulation"
        type="marker_detector.py"
        name="marker_detector"
        output="screen">
    <param name="camera_topic" value="/airsim_node/$(arg drone_name)/front_center_custom/Scene"/>
    <param name="marker_pose_topic" value="/$(arg drone_name)/marker_pose"/>
    <param name="marker_length" value="$(arg marker_length)"/>
  </node>

  <!-- Landing Controller Node -->
  <node pkg="drone_simulation"
        type="auto_landing_controller.py"
        name="auto_landing_controller"
        output="screen">
    <param name="marker_pose_topic" value="/$(arg drone_name)/marker_pose"/>
    <param name="cmd_vel_topic" value="/$(arg drone_name)/cmd_vel"/>
  </node>

  <!-- Optional RViz Visualization -->
  <group if="$(arg enable_visualization)">
    <include file="$(find drone_simulation)/launch/rviz_visualization.launch"/>
  </group>
</launch>
\end{lstlisting}

\newpage
\section{Script: \texttt{marker\_detector.py}}
\label{appendix:marker-detector}
\begin{lstlisting}[language=Python]
#!/usr/bin/env python3

import rospy
import cv2
import cv2.aruco as aruco
from cv_bridge import CvBridge
from sensor_msgs.msg import Image
from geometry_msgs.msg import PoseStamped
import numpy as np

class MarkerDetector:
    def __init__(self):
        rospy.init_node('marker_detector', anonymous=True)

        self.camera_topic = rospy.get_param('~camera_topic', '/airsim_node/drone_1/front_center_custom/Scene')
        self.marker_pub_topic = rospy.get_param('~marker_pose_topic', '/drone_1/marker_pose')
        self.marker_length = rospy.get_param('~marker_length', 0.15)  # meters

        self.bridge = CvBridge()
        self.pose_pub = rospy.Publisher(self.marker_pub_topic, PoseStamped, queue_size=10)

        self.camera_matrix = rospy.get_param('~camera_matrix', [[640, 0, 320], [0, 480, 240], [0, 0, 1]])
        self.dist_coeffs = rospy.get_param('~dist_coeffs', [0, 0, 0, 0, 0])

        self.aruco_dict = aruco.Dictionary_get(aruco.DICT_4X4_50)
        self.aruco_params = aruco.DetectorParameters_create()

        rospy.Subscriber(self.camera_topic, Image, self.image_callback)

    def image_callback(self, msg):
        try:
            frame = self.bridge.imgmsg_to_cv2(msg, desired_encoding='bgr8')
        except Exception as e:
            rospy.logerr("CV bridge error: %s", e)
            return

        gray = cv2.cvtColor(frame, cv2.COLOR_BGR2GRAY)
        corners, ids, _ = aruco.detectMarkers(gray, self.aruco_dict, parameters=self.aruco_params)

        if ids is not None:
            # Assume we detect one marker at a time for simplicity
            rvecs, tvecs, _ = aruco.estimatePoseSingleMarkers(corners, self.marker_length,
                                                              cameraMatrix=np.array(self.camera_matrix),
                                                              distCoeffs=np.array(self.dist_coeffs))
            for i in range(len(ids)):
                pose = PoseStamped()
                pose.header.stamp = rospy.Time.now()
                pose.header.frame_id = "camera"
                pose.pose.position.x = tvecs[i][0][0]
                pose.pose.position.y = tvecs[i][0][1]
                pose.pose.position.z = tvecs[i][0][2]
                # Orientation estimation from rvec is omitted for simplicity
                
                self.pose_pub.publish(pose)
                rospy.loginfo(f"Published marker pose: {pose.pose.position}")
        else:
            rospy.loginfo_throttle(5, "No marker detected.")

    def run(self):
        rospy.spin()

if __name__ == '__main__':
    try:
        detector = MarkerDetector()
        detector.run()
    except rospy.ROSInterruptException:
        pass
\end{lstlisting}

\section{\texttt{package.xml} Dependencies}
\label{appendix:package-deps}
\begin{lstlisting}[language=xml]
<package format="2">
  <name>drone_simulation</name>
  <version>0.0.1</version>
  <description>Custom drone simulation package integrating perception and control</description>
  <maintainer email="your_email@example.com">Your Name</maintainer>
  <license>MIT</license>

  <buildtool_depend>catkin</buildtool_depend>

  <build_depend>rospy</build_depend>
  <build_depend>std_msgs</build_depend>
  <build_depend>geometry_msgs</build_depend>
  <build_depend>sensor_msgs</build_depend>

  <exec_depend>rospy</exec_depend>
  <exec_depend>std_msgs</exec_depend>
  <exec_depend>geometry_msgs</exec_depend>
  <exec_depend>sensor_msgs</exec_depend>

  <exec_depend>robot_state_publisher</exec_depend>
  <exec_depend>airsim_ros_pkgs</exec_depend>

  <export>
  <!-- Other tools can request additional information be placed here -->
  </export>
</package>
\end{lstlisting}

\end{document}